\documentclass[pdflatex,sn-mathphys-ay]{sn-jnl}

\usepackage{graphicx}%
\usepackage{multirow}%
\usepackage{amsmath,amssymb,amsfonts}%
\usepackage{amsthm}%
\usepackage{mathrsfs}%
\usepackage[title]{appendix}%
\usepackage[table]{xcolor}%
\usepackage{textcomp}%
\usepackage{manyfoot}%
\usepackage{booktabs}%
\usepackage{algorithm}%
\usepackage{algorithmicx}%
\usepackage{algpseudocode}%
\usepackage{listings}%
\usepackage{caption}

\begin{document}

\title[Article Title]{Enhancing literature review with LLM and NLP methods. 
Algorithmic trading case.}


\author*[1]{\fnm{Stanisław} \sur{Łaniewski}}\email{s.laniewski@uw.edu.pl}

\author[1]{\fnm{Robert} \sur{Ślepaczuk}}\email{rslepaczuk@wne.uw.edu.pl}

\affil*[1]{\orgdiv{Department of Quantitative Finance and Machine Learning, Faculty of Economic Sciences}, \orgname{University of Warsaw}, \orgaddress{\street{Długa 44/50}, \city{Warsaw}, \postcode{00-241}, \country{Poland}}}


\abstract{This study utilizes machine learning algorithms to analyze and organize knowledge in the field of algorithmic trading. By filtering a dataset of 136 million research papers, we identified 14,342 relevant articles published between 1956 and Q1 2020. We compare traditional practices—such as keyword-based algorithms and embedding techniques—with state-of-the-art topic modeling methods that employ dimensionality reduction and clustering. This comparison allows us to assess the popularity and evolution of different approaches and themes within algorithmic trading.
We demonstrate the usefulness of Natural Language Processing (NLP) in the automatic extraction of knowledge, highlighting the new possibilities created by the latest iterations of Large Language Models (LLMs) like ChatGPT. The rationale for focusing on this topic stems from our analysis, which reveals that research articles on algorithmic trading are increasing at a faster rate than the overall number of publications. While stocks and main indices comprise more than half of all assets considered, certain asset classes, such as cryptocurrencies, exhibit a much stronger growth trend. Machine learning models have become the most popular methods in recent years. 
The study demonstrates the efficacy of LLMs in refining datasets and addressing intricate questions about the analyzed articles, such as comparing the efficiency of different models. Our research shows that by decomposing tasks into smaller components and incorporating reasoning steps, we can effectively tackle complex questions supported by case analyses. This approach contributes to a deeper understanding of algorithmic trading methodologies and underscores the potential of advanced NLP techniques in literature reviews.
}




\maketitle

\section{Introduction}\label{sec1}


The motivation for this work is to explore the extent to which automatic methods can be utilized for reviewing scientific journals, starting from the largest possible dataset, refining it with rules and machine learning to identify topics of interest, and addressing complex questions.

With the exponentially growing number of scientific journals and papers, it is difficult to keep track of how current methods evolve and change in popularity  \cite{RN165}. To organize knowledge in the field of algorithmic trading we did a thorough analysis enhanced by machine learning algorithms based on 136 million research papers from the S2ORC database, which consists of repositories such as SSRN and arXiv, Microsoft Academic Graph or international journals \cite{RN164}. 

Archives and public repositories enable the sharing of research studies at various stages of advancement, including initial preprints. As science becomes increasingly complex, interdisciplinary research involving multiple experts is on the rise. This has resulted in hundreds or thousands of papers being published each year in a specific field, making it challenging to keep track of the latest developments. 

Fortunately, advances in technology give us tools that we leverage in this study. We demonstrate how exploratory and machine-learning-enhanced analysis can be performed on a set of papers obtained by filtering one of the largest databases of research publications. LLM-assisted literature reviews offer a more dynamic and scalable solution, capturing a broader range of knowledge that manual reviews may overlook.

Our work presents a methodology that can be replicated in other fields, but we focus specifically on the case of algorithmic investment strategies. Based on automatic analysis methods such as keyword extraction and topic modeling we study a huge amount of research papers, focusing on the dynamics of selected features and how scientists apply different models. These techniques enable us to uncover insights and patterns that might have otherwise gone unnoticed, and ultimately improve our understanding of the field of algorithmic trading by providing a comprehensive list of the most successful models in these works.

\begin{center}
\bigskip
 RQ1 - Is algorithmic trading becoming a more popular topic in scientific research? How do scientific themes evolve in articles about algorithmic trading?
\bigskip
\end{center}

In addition, we identify the most popular methods and assets in this field and examine their dynamics over time. In particular, we analyze how the popularity of asset classes as a scientific topic evolves, showing that significant events are visible in aggregated statistics (e.g. rise of the popularity of cryptocurrencies). We expect that as computational power continues to increase and access to data becomes easier, shorter time horizons are studied more frequently and machine learning methods used more frequently.

\begin{center}
\bigskip
 RQ2 - How does the popularity of different asset classes, time horizons, and models studied in articles change over time?
 \bigskip
\end{center}

Finally, we want to answer which models or strategies seem to outperform the benchmark, and which parameters and hyperparameters are most important and often optimized. Such questions are difficult for keyword-based systems, as they require an understanding of concepts such as various comparison methods, the fact that models can be trained on different datasets, etc.
To address these challenges, we investigate how LLMs, including earlier generations based on Bidirectional Encoder Representations from Transformers (BERT) \cite{Bert} and both recent and state-of-the-art implementations of Generative Pre-trained Transformer (GPT) (\cite{RN204} and \cite{openai2024gpt4technicalreport}), can enhance the quantitative literature review process that currently relies on N-grams, keywords, and topic modeling. We compare different versions across time and highlight how much more insight can be distilled from full papers than abstracts.

\begin{center}
\bigskip
RQ3 - Which models outperform other models? How to optimize hyperparameters in these models?
\bigskip
\begin{itemize}
    \renewcommand\labelitemi{--}
    \item RQ3.1 - How far can we answer this question with SentenceBERT-based topic modeling?
    \item RQ3.2 - How far can we answer this question with GPT based model like ChatGPT?
\end{itemize}
\bigskip
 RQ4 - How does a version of LLM change the analysis? How much more knowledge can we get from full papers instead of abstracts?
\bigskip
\end{center}



Our key contributions are towards automation of literature review and knowledge extraction. We demonstrate how automatic methods, powered by NLP and LLMs, can replicate and enhance traditional manual literature reviews; efficiently processing, analyzing an extensive corpus of scientific literature, and ensuring that our review is comprehensive.
Secondly, through advanced topic modeling techniques, we identify patterns and trends within the field of algorithmic trading. Faster growth of papers in this topic compared to all publications indicate a growing interest and innovation in this area. We examine the dynamics of asset classes, time horizons, and models studied, highlighting how the popularity of different approaches evolves over time.
Finally, we integrate cutting-edge methodologies and utilize LLMs like ChatGPT in a novel way, comparing its performance on abstracts and full texts with questions of varying difficulty.

We structure our investigation around the research questions. Section \ref{Meth} summarise used methodology and data. To answer RQ1 and RQ2, in Section \ref{Exp} we use NLP methods such as N-grams, keywords, and Named Entity Recognition to trace the evolution of themes and trends in algorithmic trading. We employ state-of-the-art topic modeling in Section \ref{tops} to categorise the knowledge within the field and identify its trends.

We focus on selected topic of machine learning applications in trading, particularly Neural Network Trading. In Section \ref{llm_8} we compare the effectiveness of traditional keyword-based systems and embedding techniques with advanced LLMs in extracting insights about model performance and hyperparameter optimization, addressing complex questions that require a deep understanding of the content (RQ3). To answer RQ4 we assess the impact of different LLM versions on our analysis and evaluate the added value of processing full-text papers versus abstracts, highlighting the depth of insights achievable with more comprehensive data.

By addressing these questions, our work not only contributes valuable insights into the field of algorithmic trading but also presents a replicable methodology for automatic literature reviews enhanced with machine learning. We showcase how state-of-the-art NLP techniques and LLMs can transform the way researchers navigate and synthesize vast amounts of scientific knowledge, ultimately advancing the capabilities of literature analysis in the age of information overload.

\section{Literature review} 

Examples of literature reviews synthesizing available studies and technologies applied in algorithmic trading, performed using traditional manual approaches, can be seen in \cite{RNlitrev} and \cite{Joiner}. In the former article, researchers start by filtering a database to create a smaller sample and then present important quantitative statistics about the works and models, providing deeper insights into the scientific methodology by manually evaluating the papers. Meanwhile, in \cite{Joiner}, the authors follow a traditional narrative approach, explaining the data and models used in each work. 

Both studies explore the evolution of algorithmic models, their applications, and key details, which we replicate in this study. However, this conventional approach is limited by the researcher’s capacity to manually sift through an ever-expanding body of literature—a process that becomes especially challenging as fields like machine learning continue to evolve rapidly.


Recent developments in the use of LLMs for financial text mining, as discussed by \cite{SUZUKI2023103194}, where authors apply domain-specific language models in the finance domain. The main focus is comparison of performance of selected methods - namely BERT and ELECTRA - and various pre-training methods. By leveraging domain-specific language models, researchers can enhance the accuracy of NLP-based tools to extract key insights from financial reports. However, they "demonstrate no significant difference between pre-training with the financial corpus and continuous pre-training from the general language model with the financial corpus". Supporting results can be found in \cite{RN170} where ScholarBERT, a general-scientific BERT, outperforms even domain-specific embeddings, which reinforces our approach using general LLM.

We draw upon insights from \cite{Ofori}, who highlight the significant challenges and opportunities in automating systematic literature reviews (SLR) through the use of machine learning and artificial intelligence techniques such NLP and deep learning (DL). Authors present an overview of AI methods aimed at automating key stages of systematic reviews, such as search, screening, data extraction, and risk of bias assessment. Particularly relevant to our work is their exploration of NLP methods, which are crucial for handling large text datasets like ours. While \cite{Ofori} exclude LLMs "due  to  the  selection  criteria  emphasising papers with a detailed explanation of the AI methods used", they identify NLP as a key technology for classifying, retrieving, and extracting data from vast research corpora. Their insights guide our approach, where we leverage NLP to filter and employ LLMs in novel ways to analyze extensive databases of research papers, providing a more systematic and comprehensive review of the selected literature.

Integrating automation into SLR can significantly improve efficiency by employing intelligent filtering mechanisms to reduce the volume of literature under consideration. An example of SLR with smart filtering in the topic of algorithmic trading can be found in \cite{GUNNARSSON2024103221}. However, after refining the scope down to 32 papers, they did the review manually, which contrasts with our novel approach to apply NLP and LLM to extract that knowledge.

Another approach is presented in \cite{RNCachola}, where the summarisation model for scientific papers is fine-tuned by using training datasets created by experts. This can used for enhancing LLMs ability to condense and extract meaningful content from complex research articles.

Filtering and synthesising knowledge yields a refined subset of documents that is well-suited for expert annotation and subsequent analysis. For instance, \cite{RN195}  applied keyword-based filtering to initially reduce their dataset to 3,740 papers. They then manually annotated these papers to create a training set for two supervised machine learning models—a Support Vector Machine (SVM) and a Convolutional Neural Network (CNN)—designed to classify the papers into two distinct categories.

In a similar study, \cite{RN190} conducted an SLR focusing on feature selection methods for text classification, analyzing 175 papers published between 2013 and 2020. They performed temporal analyses and visualizations to assess the evolving popularity of various feature selection techniques within the scientific community. Their methodology also relied on keyword filtering and manual expert analysis.

However, \cite{RN188} caution that such approaches may lack robustness and sustainability. They argue that maintaining the necessary software tools can be time-consuming and costly. Moreover, they highlight the unique challenges associated with medical SLRs, suggesting that while machine learning models can screen out non-randomized controlled trials, alternative applications of machine learning may offer more value in the context of literature reviews.

To address some of these limitations, \cite{RN202} enhanced the manual literature review process by incorporating CiteSpace for automated citation visualization. Their study included analyses of journals and geographical distribution of research, topic-based document clustering, and the use of word clouds and keyword burst analysis to identify emerging trends in the literature.

For topic modeling we follow BERTopic introduced in \cite{RN200}, which has been successfully used in research, e.g. \cite{RN203}. Our work is at the forefront of enhancing literature review methodologies using GPT models (\cite{RN204}), comparing the performance of different model versions and the use of abstracts versus full-text analysis. Previous attempts like \cite{RN194} focused on writing new papers, while we use it for knowledge synthesis and answering intricate questions that keywords-based methods cannot. We follow findings of how to cope with GPT issues such as hallucination \cite{RNli2023halu} or change in performance over time \cite{tu2024chatlog}.

\section{Methodology}\label{Meth}
The methodology used in the paper can be summarised as follows.

To refine our dataset, we employed a comprehensive search strategy and various filtering techniques based on keywords, topics, expert knowledge, and journals. Then we evaluated multiple embedding methods, including word2vec and universal-sentence-encoder, before selecting sentence BERT as the optimal approach for our problem. To perform topic modeling, we utilized state-of-the-art algorithms, such as BERTopic \cite{RN200}, which involved dimension reduction using Uniform Manifold Approximation and Projection (UMAP) \cite{RN198} and clustering with Hierarchical Density-Based Spatial Clustering of Applications with Noise (HDBSCAN) \cite{RN199}.

We curated the outcomes of topic modeling algorithms to find the major themes and analyze which areas of research are growing most rapidly.

We calculated embeddings for our research questions and identified topics with the closest cosine distance. We subsequently validated our results by scrutinizing the papers using both expert knowledge and a GPT-based model (ChatGPT). Finally, we demonstrate an automated procedure involving LLMs, which reproduces the researchers' process during literature reviews when searching for properties of the papers (e.g. the results of model comparisons).

To illustrate our findings, we present an analysis that includes a statistical overview and supporting visualizations that highlight distinctions in papers, including the algorithms employed, the markets, and the main subjects of study. We also examine these differences across various dimensions such as time and popularity.

\subsection{Data selection}

To ensure we review a broad range of relevant research, we began by identifying the most suitable database. After careful consideration, we selected the S20RC database \cite{RN164}, which is a huge corpus of over 136M scientific papers enriched with citation data derived from Semantic Scholar, a research tool developed at the Allen Institute for AI. They span over 70 years with the last entries from April 2020.

We designed a schema to extract a relevant database (corpus) of documents. Our approach could be replicated for any research topic, but we focused on algorithmic trading, which required a smart filtering process. We considered that essential research, models, or findings could be outside the scope of regular economic journals or be interdisciplinary. 

Our filtering reduced the corpus to 16 197 documents. We removed from further analysis those for which we could not fetch an abstract, ending up with 14,342 documents. Table \ref{tab:regf} presents the statistics for each keyword used in the process. Investment strategies were found to be the most popular, with almost 3k occurrences in titles and 10k in abstracts. The other keywords were mentioned over 2k in the title and 4k in the abstract.

\begin{table}
\centering
\caption{Frequency of keywords}
\label{tab:regf}
\rowcolors{2}{gray!25}{white}
\begin{tabular}{l|c|c|c}
\hline
\rowcolor{gray!50}  
\textbf{Regular expression} & \textbf{Abstract} & \textbf{Title} & \textbf{Both} \\ 
\hline
Algo(rithmic)* trading & 615 & 390 & 841 \\
Investment strateg. & 9473 & 2921 & 11362 \\
Vola(tility)* trading & 86 & 54 & 129 \\
High.frequency trading & 832 & 719 & 1248 \\
Investment system. & 870 & 174 & 963 \\
Benchmark strateg. & 170 & 7 & 177 \\
Pair.trading & 67 & 35 & 85 \\
Momentum (trading $\vert$ strateg.) & 1074 & 461 & 1315 \\
Contrarian (trading $\vert$ strateg.) & 380 & 169 & 477 \\
\hline
\rowcolor{gray!50}
\textbf{SUM} & \textbf{13567} & \textbf{4930} & \textbf{16597} \\
\hline
\end{tabular}
\end{table}

\section{Exploratory analysis}\label{Exp}
To address our research question about the popularity of algorithmic trading strategies and methods over time, we conducted analyses of the dataset, including publication dates, citation data, and keyword-based and topic-modeling methods.

Furthermore, we preprocessed the collected documents by removing English stopwords, lemmatizing the words, and tokenizing the texts. We also calculated descriptive statistics to provide a detailed overview of the documents we collected. 

To understand the corpus more thoroughly, we generated word clouds and found N-grams and noun chunks. The generated world cloud confirms we have captured relevant articles from the targeted domain. Analysis of N-grams revealed the most popular concepts such as efficient market hypotheses, limit order book signals, time series momentum, and models such as the Fama French factor model and capital asset pricing model (CAPM). Additionally, we used Named Entity Recognition algorithms to identify the most commonly studied markets and countries in the scientific literature related to our selected topics.

\subsection{Time horizon and top asset classes}
\begin{figure}
\centerline{\includegraphics[width=0.72\textwidth]{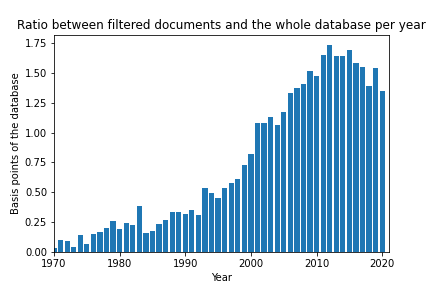}}
\caption{\label{fig:bps}How many basis points (0.0001) of whole S2ORC is in our dataset}  
\end{figure}

We want to highlight three findings: first, the increasing popularity of algorithmic investment strategies by showing how it is becoming a greater part of the total database in Figure \ref{fig:bps}.

\begin{figure}
\centerline{\includegraphics[width=0.95\textwidth]{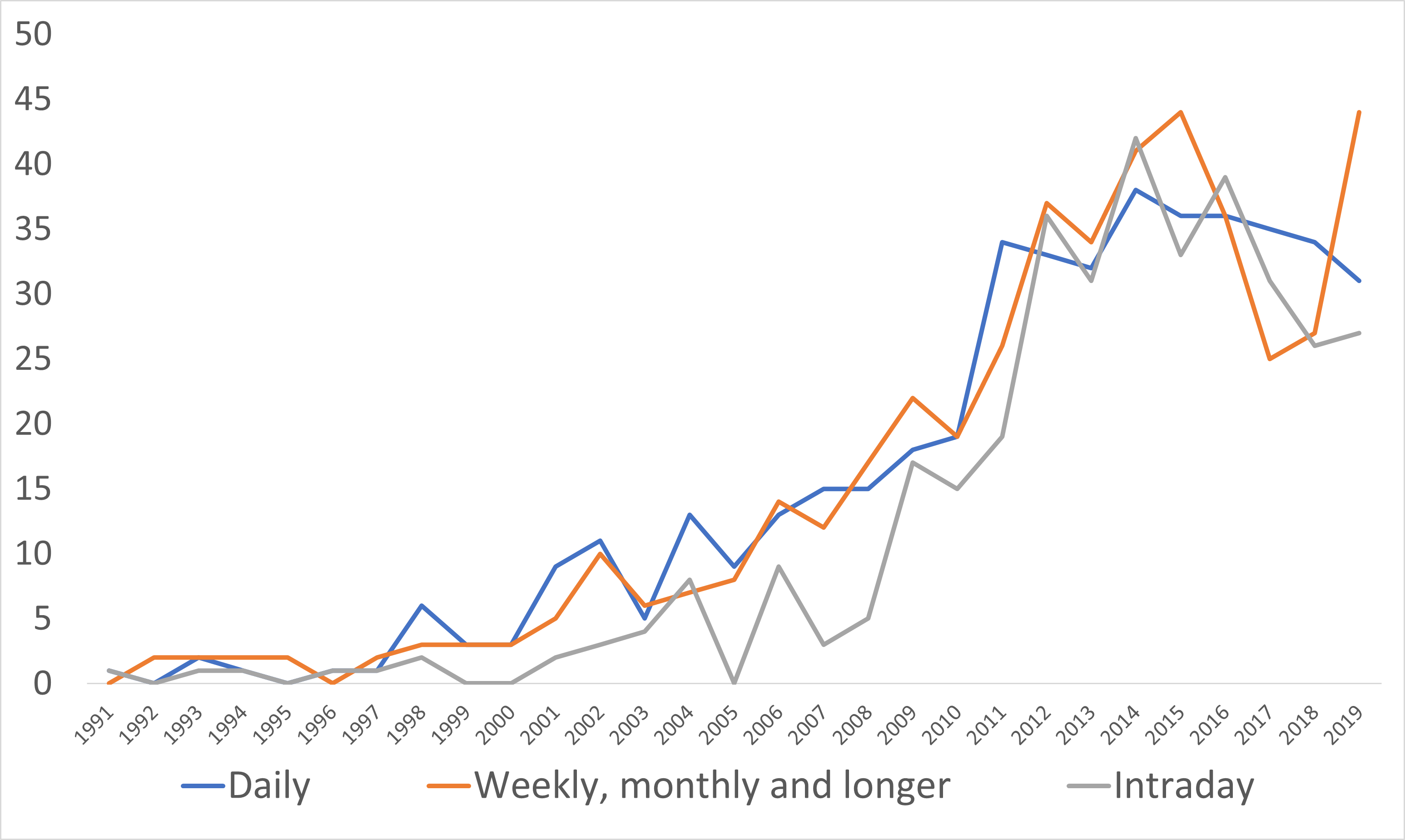}}
\caption{\label{fig:timehor}Popularity of time horizons over years based on keywords in the context of data sampling}
\end{figure}

Second, the majority of publications deal with daily and monthly data. In Figure \ref{fig:timehor} we plot the frequency of each time horizon, which we defined by analyzing keywords such as "daily", or "5-minute", or "monthly" in the context of data or training periods. The evaluation based on sampling from results and checking manually gave an 80\% positive rate.

\begin{figure}
\centerline{\includegraphics[width=0.95\textwidth]{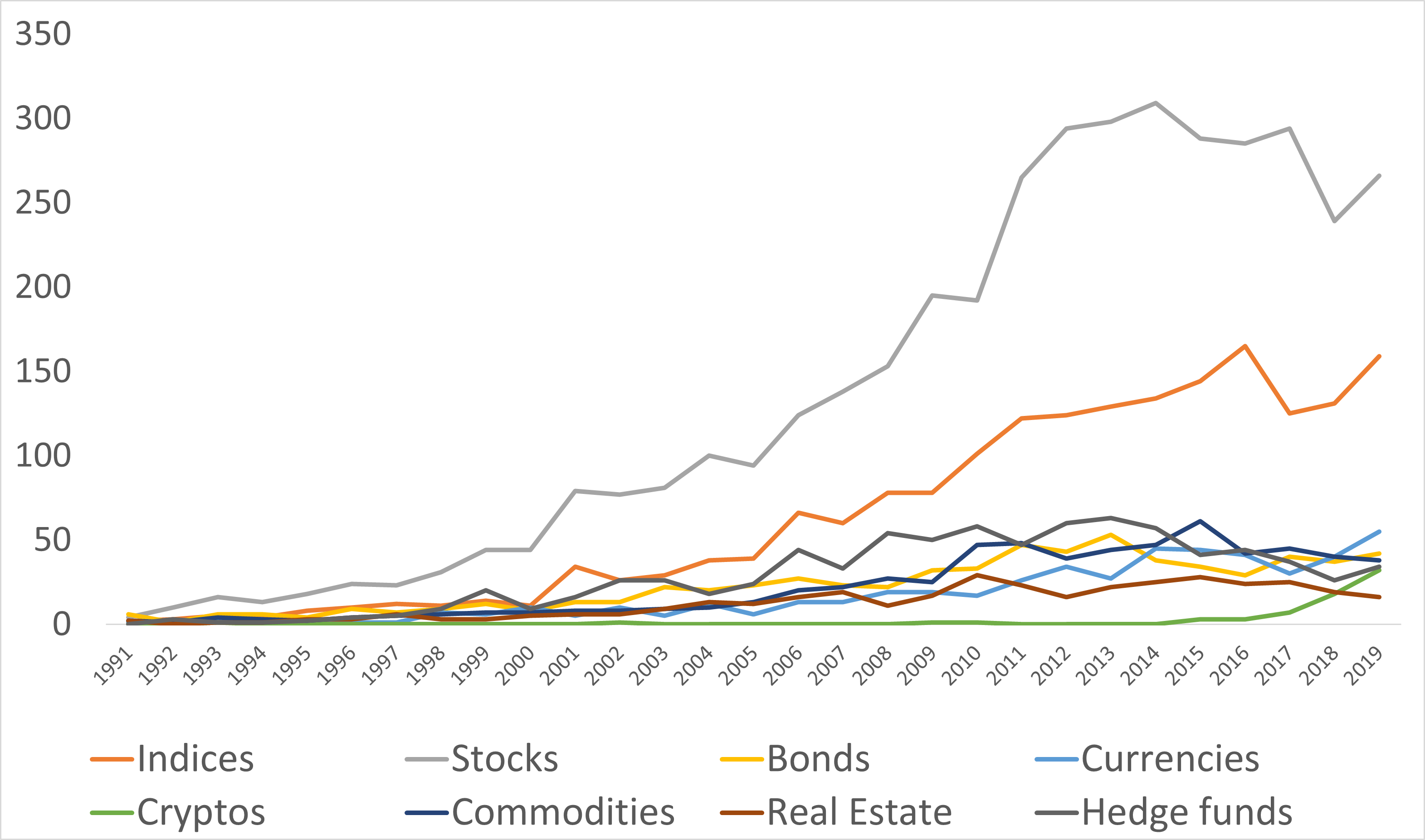}}
\caption{\label{fig:ac}Popularity of asset classes in time}
\end{figure}

By applying the same approach (Figure \ref{fig:ac}), we also found that researchers tend to focus on stocks, their indices, and derivatives, with more than half of the papers covering these topics. Cryptocurrencies have only recently become a topic of scientific interest, we also noticed increased interest in commodities around the time of the 2014-2016 oil crisis, which saw a 70 percent price drop.

\subsection{Top methods used for modelling}
To answer the question of the increasing popularity of machine learning-based methods in recent years, we aggregated them and compared them to linear models and time series. Although linear models account for more than half of all methods considered in the database, we examine the trends of different model families over time. We do this by using regular expressions to search for specific model names and grouping them into three categories: linear, time series, and machine learning (Figure \ref{fig:modd}). We then plot the number of papers per year that mention each category of models (full regex in Appendix \ref{sub:reg}).

\begin{figure}
\centerline{\includegraphics[width=0.95\textwidth]{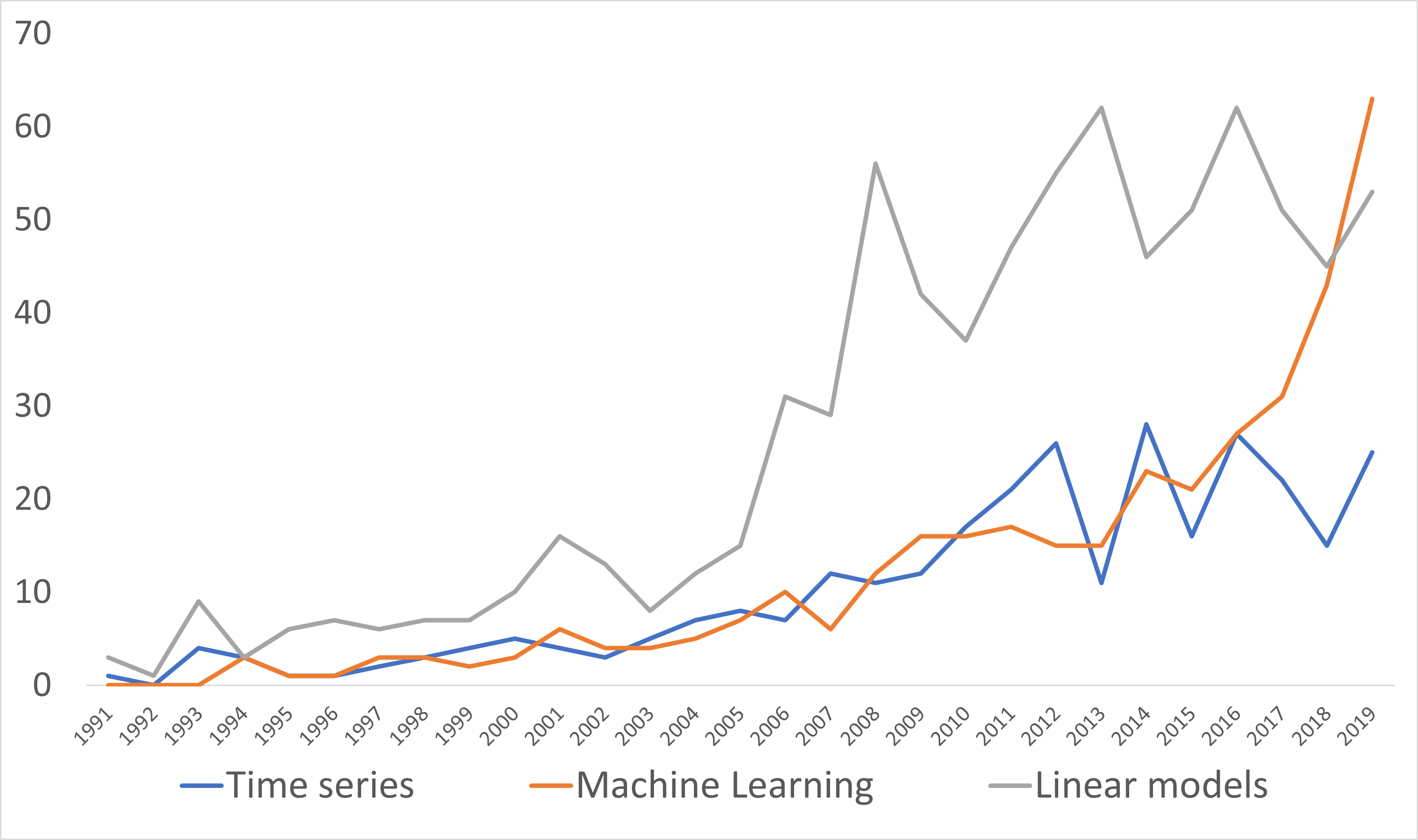}}
\caption{\label{fig:modd}Popularity of models classes in time}
\end{figure}

Our analysis shows that machine learning methods are rapidly gaining popularity in algorithmic trading research, especially since 2015. In 2019 machine learning methods surpassed linear models in popularity for the first time in history. This supports trends found by previous researchers, e.g. \cite{RNlitrev}. The neural network is the most researched system from a machine learning environment. On the other hand, the use of time series modeling appears to be losing traction, with a decreasing ratio of papers incorporating them in the algorithmic trading scope.

\section{Topics}\label{tops}

\subsection{Procedure}
To delve deeper into the underlying topics of the research papers, we recognise the need to augment our analysis with more advanced techniques. The statistical-based methods employed thus far have provided valuable insights, but to gain a more nuanced understanding, we require the ability to comprehend language, identify similarities between words and sentences, and extract meaningful summaries from the texts. By doing so, we can uncover the topics that are considered crucial by the scientific community.

To do so we apply three various embeddings word2vec, BERT, and Universal Sentence Encoder - to better understand the language, find similarities between words and sentences, and summarise the texts. While word2vec has been successfully used in previous research, we found that its lack of sentence interpretation could be a potential flaw in our analysis. After testing various sentence transformer-based methods, we ultimately chose the 384-dimensional all-MiniLM-L6-v2 model for its superior performance while maintaining a small size, based on \cite{RN197}.

To reduce the dimensionality of our embedded documents, we use UMAP (\cite{RN198}). It is a non-linear dimension reduction algorithm that combines aspects of principal component analysis (PCA) and t-distributed stochastic neighbor embedding (t-SNE). By using UMAP, we aimed to preserve the essential global structures of the documents, making it easier to identify similar topics.

After applying UMAP, we further grouped the documents using the HDBSCAN algorithm \cite{RN199}, as it was capable of detecting clusters of different densities and able to hand outliers. Therefore documents that had similar embeddings, such as those of the same topic or containing significant word overlap, end up grouped together based on their lower-dimensional representations.

Following the implementation of BERTopic \cite{RN200} we used the class version of term frequency–inverse document frequency (TF-IDF). The documents which fall into the same cluster create a topic.

Finally, we curated outcomes of topic modeling algorithms to identify common themes and determine which areas of research are growing most rapidly. We used a technique of merging smaller topics with the closest thematically larger ones based on the shortest Euclidean distance in the lower dimensional space. This allowed us to reduce the number of topics to 20. To validate our results, we sampled 50 documents and manually checked them. Lastly, we prompt ChatGPT to generate a 3-word title for each topic based on the top 10 words and scores from the TF-IDF table.

\subsection{Analysis}

\begin{figure}
\centerline{\includegraphics[width=0.98\textwidth]{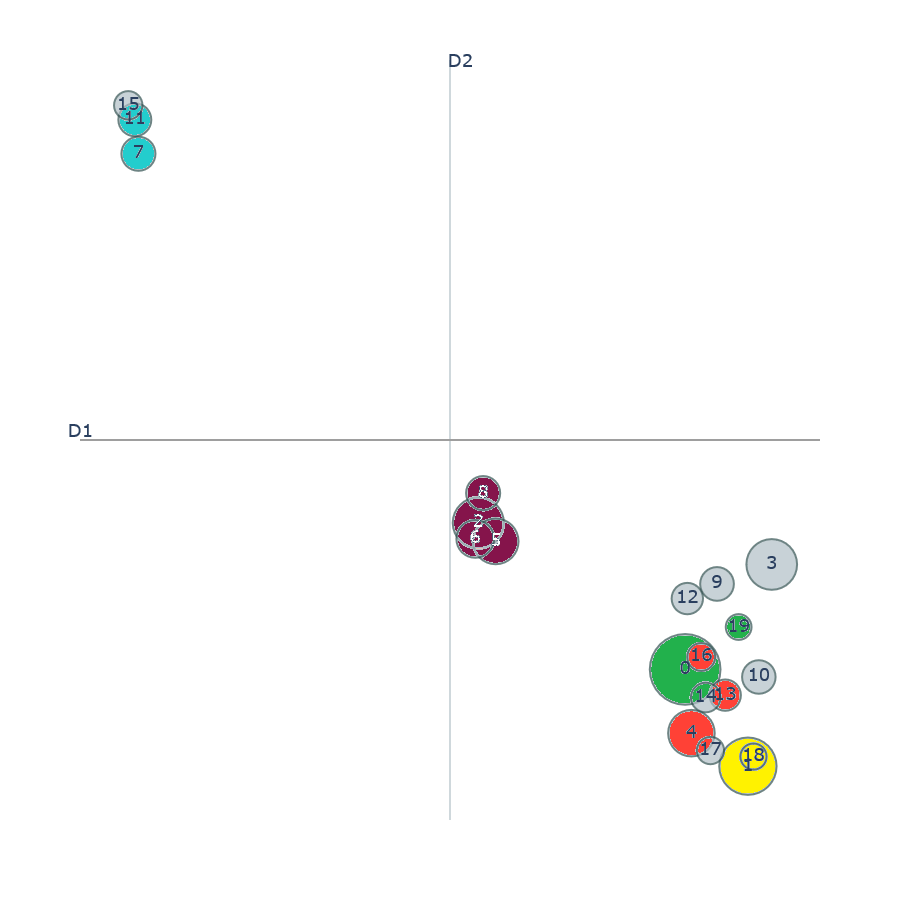}}
\caption{\label{fig:Topicc}Topic clusters. The colors are shared with \ref{fig:Hierarchical}, where the most similar topics are grouped together.}
\end{figure}

We identified three distinct clusters: one major group and two smaller (Figure \ref{fig:Topicc}). The middle cluster is centered on strategic investments, such as those in transportation, the military, and electricity, while the top-left cluster focuses on education, agriculture, foreign investment, and development. The main group in the bottom-right is about investment strategies; we notice some sub-clusters about the pension system (upper part of the group), the main part consisting of various strategies, and in the bottom right optimal investments for longer periods (from portfolio manager and insurer perspective).

\begin{figure}
\centerline{\includegraphics[width=0.98\textwidth]{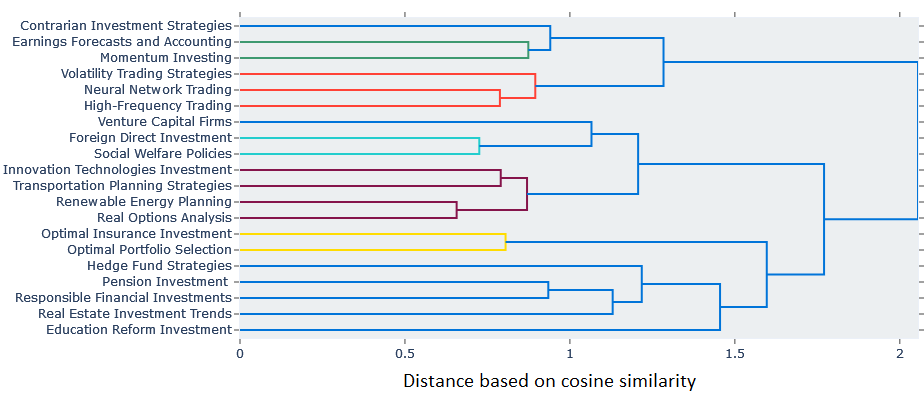}}
\caption{\label{fig:Hierarchical}Hierarchical clustering. Topics which are similar will be merged first, based on cosine similarity.}
\end{figure}

We can also notice that topic 16 (Neural Network Trading) is interestingly first matched with topic 4 (High-Frequency Trading, HFT) and 13 (Volatility) rather than general group 0, which matches with 19 (Figure \ref{fig:Hierarchical}). This suggests that topic 16 has stronger connections with the specialized areas covered by topics 4 and 13.

Topics that are close to each other based on our model will be grouped first. Comparing to  \ref{fig:Topicc}, we notice that indeed most often topics that were clustered together are grouped first, e.g. 7: Foreign Direct Investment and 11: Social Welfare Policies, or two pairs 2 \& 5 and 6 \& 8. (2: Renewable Energy Planning, 5: Real Options Analysis, 6: Innovation Technologies Investment, 8: Transportation Planning Strategies).

\begin{figure}
\centerline{\includegraphics[width=1.0\textwidth]{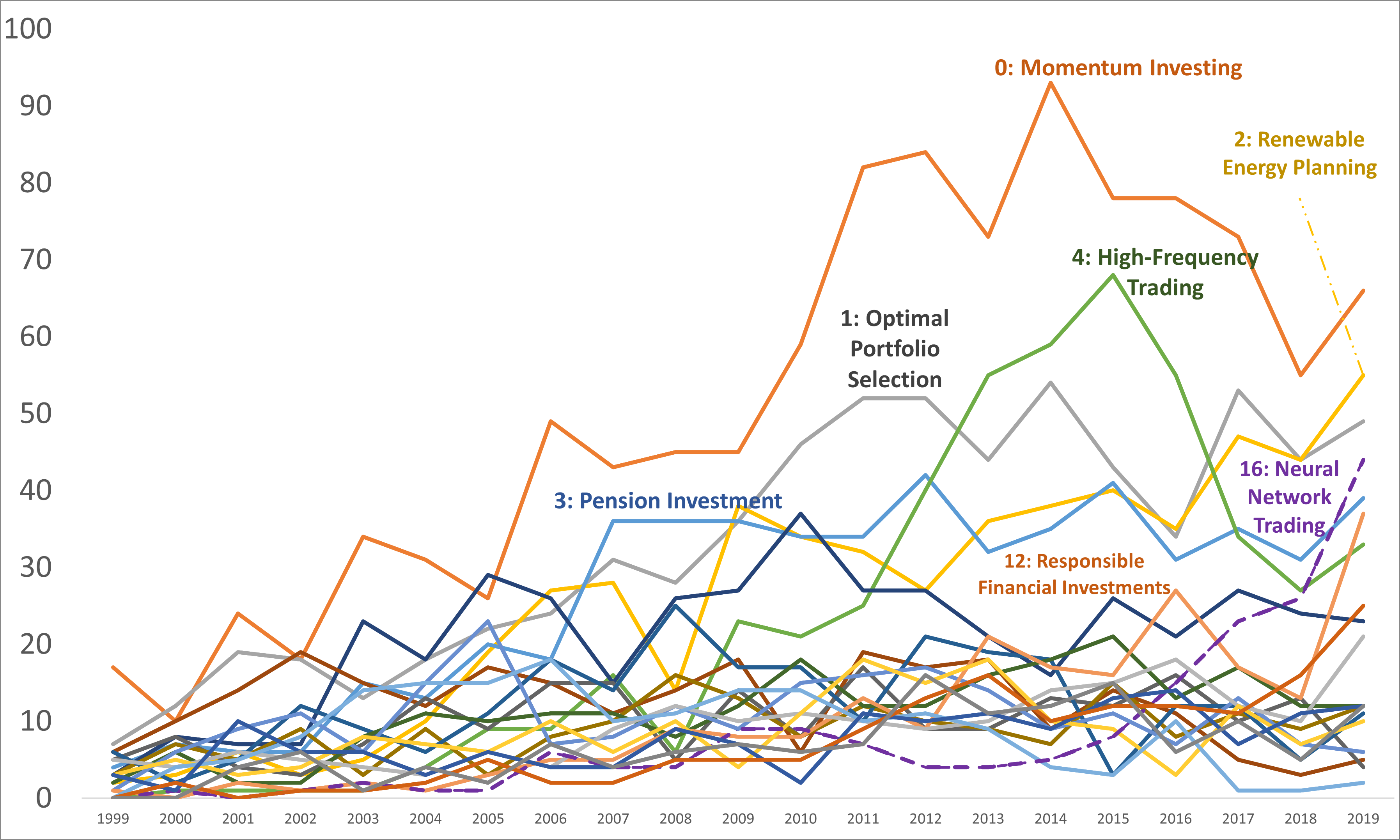}}
\caption{\label{fig:word}The trend of 20 main topics in the last 20 years with labels for the top 7 topics in 2019}
\end{figure}

In Figure \ref{fig:word} we notice that trends change over time, for example, topic 4 (HFT) experienced bursts of popularity in response to events such as the first flash crash. On the other hand, topic 16 (Neural Network Trading) has become one of the fastest-growing areas of research in this field in recent years.

\subsection{Neural Network Trading}
To analyze the topic that compares different models used in algorithmic trading, we started by identifying relevant keywords and queries, such as "model outperforms", "hyperparameter optimization", "learning rate", "comparing models", and specific method-related terms like "recurrent neural network", "long short-term memory/LSTM", and "reinforcement learning". For each query we created an embedding, compared them with the topic embeddings, and identified the most similar topics by calculating the distance based on cosine similarity values (Table \ref{tab:que}, the higher Simil. the better).

\begin{table}
\centering
\caption{Top topics for selected queries}
\label{tab:que}
\rowcolors{2}{gray!25}{white}  
\begin{tabular}{|c|c|c|c|c|c|c|}
\hline
\rowcolor{gray!50}  
\textbf{Method} & \textbf{Value} & \textbf{1\textsuperscript{st}} & \textbf{2\textsuperscript{nd}} & \textbf{3\textsuperscript{rd}} & \textbf{4\textsuperscript{th}} & \textbf{5\textsuperscript{th}} \\ 
\hline
Model            & Topic & 16    & 5     & 1     & -1    & 2     \\
\cellcolor{white}Outperforms      & Simil. & 0.43  & 0.30  & 0.30  & 0.28  & 0.28  \\ 
\hline
Learning Rate    & Topic & 16    & 15    & 2     & 1     & 5     \\
\cellcolor{white}                 & Simil. & 0.42  & 0.31  & 0.28  & 0.27  & 0.26  \\ 
\hline
Hyperparameter   & Topic & 16    & 1     & 2     & 5     & 13    \\
\cellcolor{white}Optimization     & Simil. & 0.34  & 0.23  & 0.22  & 0.20  & 0.19  \\ 
\hline
Comparing        & Topic & 16    & 13    & 5     & 2     & 6     \\
\cellcolor{white}Models           & Simil. & 0.33  & 0.24  & 0.23  & 0.23  & 0.21  \\ 
\hline
LSTM             & Topic & 16    & 8     & 19    & 13    & 18    \\
\cellcolor{white}                 & Simil. & 0.50  & 0.26  & 0.26  & 0.25  & 0.25  \\ 
\hline
Recurrent        & Topic & 16    & 1     & 2     & 18    & -1    \\
\cellcolor{white}Neural Network   & Simil. & 0.58  & 0.30  & 0.28  & 0.28  & 0.27  \\ 
\hline
Reinforcement    & Topic & 16    & 2     & 5     & 1     & 6     \\
\cellcolor{white}Learning         & Simil. & 0.60  & 0.51  & 0.46  & 0.44  & 0.43  \\ 
\hline
\end{tabular}
\end{table}



For each of our queries, topic 16 (Neural Network Trading) gets the highest cosine similarity. The 10 top words of this topic include trading, prediction/forecasting, neural network, stock, and machine learning.
  
In the limited domain of 176 research papers, we conducted a detailed analysis to answer our research questions: what assets and venues are most frequently used, how are they tested (models), and which techniques perform better (Table \ref{tab:how2} and Tables \ref{tab:what}, \ref{tab:how} in Appendix). We use keyword- and LLM-based methods and validate them by asking experts to read through the abstracts as well.


\begin{table}
\centering
\caption{Top 5 techniques used in topic 16 (Neural Network Trading)}
\label{tab:how2}
\rowcolors{2}{white}{gray!25}  
\begin{tabular}{|l|c|}
\hline
\rowcolor{gray!50}
\textbf{Model} & \textbf{Count} \\ \hline
\rowcolor{white}Neural Network (NN) & 80 \\
\rowcolor{gray!25}Imitation Learning, Reinforcement Learning, Q-Learning/Network, & \\
\rowcolor{gray!25}Actor-Critique, A3C & 62 \\
Machine Learning & 48 \\
Data Mining, Rough Set, Fuzzy & 37 \\
\rowcolor{white!25}Technical Analysis (TA), Technical Indicator, Oscillator, & \\
\rowcolor{white!25}Moving average convergence/divergence (MACD)  & 36  \\
\hline
\end{tabular}
\end{table}

As expected, this topic is dominated by neural networks and reinforcement learning. Nonetheless, other methods such as rough sets for data mining, support vector regression, support vector machines, and various mapping or pattern searching algorithms are also prominent. Several models incorporate machine learning-based feature creation, particularly based on technical analysis. Linear models and Buy-and-Hold strategies are often used as benchmarks, although not always since some B\&H strategies are based on classification models for market entry and exit.

For model comparison or hyperparameter optimization (HPO), unfortunately, frequently there was minimal detailed information beyond statements such as 'Model X is compared 
 with model Y'. Sometimes they were just contrasted with simple benchmarks or some other strategy with inconclusive results (not mentioned in the abstract, nor the comparison method). Additional keyword frequency listed in Table \ref{tab:kilka22} suggests that a more sophisticated model is required for such complex questions.

\begin{minipage}{0.33\textwidth} 
\centering
\captionof{table}{How often were the models compared?}
\label{tab:kilka22}
\rowcolors{1}{white}{gray!25}  
\begin{tabular}{c|c}
\hline
\rule{0pt}{12pt}
Compare & 51 \\
Accuracy & 39 \\
Outperform & 29 \\
Benchmark & 18 \\
Sharpe & 5 \\
Precision, Recall, F1 & 2 \\
\hline
\end{tabular}
\end{minipage}%
\hfill 
\begin{minipage}{0.67\textwidth} 
\centering
\captionof{table}{What does ChatGPT tell us?}
\label{tab:LLM}
\begin{tabular}{l|c|c}
\hline
\rule{0pt}{12pt}
 & No model & Comparing  \\
 & comparison & models \\
\rule{0pt}{12pt}
No HPO & 64 & 48 \\
With HPO & 35 & 29 \\
\hline
\end{tabular}
\end{minipage}

\section{Analysis with LLM}\label{llm_8}

To answer RQ3 \& RQ4 and to evaluate the efficiency of LLM, we test two ChatGPT models, specifically ChatGPT 3.5 (23.03.23) and ChatGPT-4o (01.06.24), on the selected subset of papers, namely the ones labeled in topic modeling as topic 16 - Neural Network Trading.

\subsection{Comparing GPT versions on abstracts}

We employed both ChatGPT versions 3.5 and 4o to answer RQ3 regarding comparing models and HPO. We designed a prompt that asks if each abstract contains two aspects: a comparison of different models or methods used and hyperparameter optimization (Table \ref{tab:LLM}). We required each answer to be summarised with a yes/no response. To validate the results, we manually evaluated the abstracts and checked the longer answers provided by the LLM.

The 4o ChatGPT reveals an increase in the number of papers identified as comparing models and performing HPO. The overall number of papers without model comparison decreased by 17, illustrating the effectiveness of the 4o approach in uncovering methodological details.

\begin{table}[h!]
    \centering
    \caption{Confusion Matrix Comparing LLM versions (3.5 Turbo to 4o)}
    \rowcolors{3}{gray!25}{white}
    \begin{tabular}{llrrr}
        \rowcolor{gray!50}
        \toprule
        \cellcolor{white} & \cellcolor{white} & \textbf{4o Abstracts} & \textbf{3.5 Abstracts} & \textbf{Difference} \\
        \rowcolor{gray!50} \multirow{-2}{*}{\cellcolor{white}\textbf{HPO}} & \multirow{-2}{*}{\cellcolor{white}\textbf{Category}} & \textbf{Count} & \textbf{Count} & \textbf{(4o - Abstracts)} \\
        \midrule
        \cellcolor{white} & No model comparison & 47 & 64 & -17 \\ \multirow{-2}{*}{\cellcolor{white}No HPO}
        & Comparing models & 98 & 48  & 50 \\
        \midrule
        \cellcolor{white} & No model comparison & 6 & 35 & -29 \\ \multirow{-2}{*}{\cellcolor{white}With HPO}
        & Comparing models & 25 & 29 & -4 \\
        \midrule
        \cellcolor{white} & \textbf{Total Sum} & 176 & 176 & \textbf{0} \\
        \bottomrule
    \end{tabular}
    \label{tab:confusion-matrix-abstracts-4o}
\end{table}

In 3.5, model comparison was not observed in 99 abstracts, whereas in 4o, it was noted in only 53 abstracts, marking a difference of 46 papers. A significant portion of this difference can be attributed to the classification of abstracts as comparing models without HPO: 98 abstracts in 4o compared to 48 abstracts in 3.5, accounting for a difference of 50 papers. Additionally, 4o adopted a more stringent criterion for HPO, identifying only 31 abstracts as employing such methods, compared to 64 abstracts in 3.5 (difference of 33 papers). 

\subsection{Full text analysis}

From the 176 articles on the topic of Neural Network Trading, we accessed 153 full papers and removed 7 biggest files (books), ending up with 146 full texts for analysis.

Furthermore, since we analyze full papers now, we asked more elaborate questions. There are 3 new questions added to model comparison and HPO, namely frequency of data used, loss function used, and what was chosen as the best model. We expect two answers for each question - one with yes/no, the other with the explanation provided for each question.
\begin{verbatim}
    If there is a comparison of different models or methods used.
    If there is hyperparameter optimization.
    The frequency of data used.
    The loss function used.
    The best model (chosen in comparison).    
\end{verbatim}

\subsubsection{Comparing to Abstracts}

Despite having 30 fewer full texts than abstracts, the LLM was able to find snippets where researchers compared models or performed HPO, leading to a significant increase of 59 affirmative answers to both questions. Additionally, the overall number of papers without model comparison decreased by 42, illustrating the effectiveness of full-text analysis in uncovering methodological details. 

The significant increase in the detection of model comparisons and HPO highlights the necessity of full-text analysis for comprehensive research reviews. This study demonstrates that intricate methodological nuances are often embedded deeper in the papers, which can be effectively uncovered using advanced language models.

\begin{table}[h!]
    \centering
    \caption{Confusion Matrix Comparing Full Texts and Abstracts}
    \rowcolors{3}{gray!25}{white}
    \begin{tabular}{llrrr}
        \rowcolor{gray!50}
        \toprule
        \cellcolor{white} & \cellcolor{white} & \textbf{Full Texts} & \textbf{Abstracts (4o)} & \textbf{Difference} \\ 
        \rowcolor{gray!50} \multirow{-2}{*}{\cellcolor{white}\textbf{HPO}} & \multirow{-2}{*}{\cellcolor{white}\textbf{Category}} & \textbf{Count} & \textbf{Count} & \textbf{(Full Texts - 4o)} \\
        \midrule
        \cellcolor{white} & No model comparison & 6 & 47 & -41 \\ \multirow{-2}{*}{\cellcolor{white}No HPO}
        & Comparing models & 51 & 98 & -47 \\
        \midrule
        \cellcolor{white} & No model comparison & 5 & 6 & -1 \\ \multirow{-2}{*}{\cellcolor{white}With HPO}
        & Comparing models & 84 & 25 & 59 \\
        \midrule
        \cellcolor{white} & \textbf{Total Sum} & 146 & 176 & \textbf{-30} \\
        \bottomrule
    \end{tabular}
    \label{tab:confusion-matrix-full-texts-4o}
\end{table}

We notice increased detection of model comparison and HPO (comparing models and performing HPO rose from 123 and 31 in abstracts to 135 and 89 in full texts respectively) and reduction in papers without methodological details (not comparing models, not performing HPO fall from 53 and 145 in abstracts to 11 and 57 in full texts respectively).

\subsubsection{Time intervals}

The full-text analysis provides us with more accurate information about the frequency of data. We defined the bins by taking a list of unique answers (\ref{App:Czasy}) and grouping them manually. 

\begin{table}[h!]
    \centering
    \caption{Frequency of Data Used}
    \rowcolors{2}{gray!25}{white}  
    \begin{tabular}{lrrrr}
    \rowcolor{gray!50}
        \toprule
        & \textbf{Intraday} & \textbf{Daily} & \textbf{Longer} & \textbf{Not specified} \\
        \midrule
        \textbf{Count} & 37 & 73 & 24 & 12 \\
        \midrule
        \textbf{Regex on Abstracts} & 15 & 16 & 8 & 119 \\
        \bottomrule
    \end{tabular}

    \label{tab:frequency-data-used}
\end{table}

\subsubsection{Loss functions}
The majority of the loss functions fall into the "Other/Unspecified" category, indicating a variety of less commonly named or unique loss functions. The most commonly specified loss function group is MSE-related, followed by Cross-entropy-related.

\begin{table}[h!]
    \centering
    \caption{Summary of Loss Function Instances}
    \rowcolors{2}{gray!25}{white}  
    \begin{tabular}{lr}
    \rowcolor{gray!50}
        \toprule
        \textbf{Loss Function Group} & \textbf{Instances} \\
        \midrule
        MSE Related & 28 \\
        Cross-Entropy Related & 13 \\
        Other Common Loss Functions & 11 \\
        Specialized/Custom Loss Functions & 11 \\
        RMSE Related & 8 \\
        Sharpe Ratio Related & 4 \\
        MAPE Related & 1 \\
        Other/Unspecified & 69 \\
        \bottomrule
    \end{tabular}
    \label{tab:loss-function-summary}
\end{table}
To present the results, we grouped the loss functions based on expert knowledge \ref{Lossy}.

\subsubsection{Best models}
Here is the summary based on NLP and lazy ChatGPT (that is, the one that uses Python to analyze data instead of reading manually, as it's longer than its context).
\begin{table}[h!]
    \centering
    \caption{Summary of Best Model Categories based on NLP and ChatGPT}
\rowcolors{2}{gray!25}{white}  
    \begin{tabular}{lr}
    \rowcolor{gray!50}
        \toprule
        \textbf{Model Category} & \textbf{Count} \\
        \midrule
        Neural Networks & 25 \\
        Traditional Statistical Models & 13 \\
        Recurrent Neural Networks & 12 \\
        Reinforcement Learning & 5 \\
        Self-Organizing Maps (SOMs) & 4 \\
        Ensemble Methods & 3 \\
        Fuzzy Logic Models & 3 \\
        Other/Unspecified & 70 \\
        \bottomrule
    \end{tabular}

    \label{tab:best-model-summary}
\end{table}

\ref{tab:detailed-best-model-summary} is the detailed categorization of models that reflects the thorough analysis performed by prompting LLM with each answer and its corresponding elaboration. This method of batching ensures that the LLM meticulously 'reads' the elaboration, leveraging its capabilities to provide accurate and insightful classifications.

Not only does this result in a more precise classification of models (with the "Other/Unspecified" category dropping from 70 to 20), but it also captures more categories and nuances, such as creating a distinct topic for deep learning models. This enhanced granularity in classification demonstrates the LLM's capability to discern subtle differences and provide a comprehensive overview of the diverse range of models used in the studies.

\begin{table}[h!]
    \centering
    \caption{Detailed Summary of Best Model Categories}
\rowcolors{2}{gray!25}{white}  
    \begin{tabular}{lr}
    \rowcolor{gray!50}
        \toprule
        \textbf{Model Category} & \textbf{Count} \\
        \midrule
        Deep Learning Models & 25 \\
        Traditional Statistical Models (including 6 Trees) & 21 \\
        Neural Networks & 19 \\
        Recurrent Neural Networks and extensions & 13 \\
        Reinforcement Learning & 11 \\
        Ensemble Methods and Hybrid Models & 11 \\
        Specialised Models & 11 \\
        Support Vector Machine Models& 8 \\
        Rough sets & 7 \\
        Not applicable/Unspecified & 20 \\
        \bottomrule
    \end{tabular}

    \label{tab:detailed-best-model-summary}
\end{table}

The full list can be found in \ref{Modele}.

\subsection{Issues}

\subsubsection{LLM Laziness}

First of all, the answering scheme is different than one would imagine AI to use. By default, ChatGPT does not read and understand the papers. Instead, it uses regex and NLP methods to answer each question. When prompted, it even provided us with the Python code it used for analysis \ref{lazyGPT}. As expected, it is suspect to simple false positives (the word 'compare' is used in different contexts) or to omitting keywords (not provided in the short list of options).

For example, \ref{tab:yes_no_counts} is the initial analysis provided for each question, as it simply treated it all as one batch and ran a regex analysis on it. It performed surprisingly well - the words selected in regex produced quite accurate results for the number of papers with model comparison (138 compared to 135 based on LLM full-text analysis, \ref{tab:confusion-matrix-full-texts-4o}) and for HPO (96 compared to 89). In data frequency, question it had problems capturing intraday horizons, thus predicting 107 papers to state frequency of dataset used, compared to 134 in LLM full paper analysis \ref{tab:frequency-data-used}. Again it did well in the loss function, stating that 65 are unspecified, compared to 69 in \ref{tab:loss-function-summary}, while the choice of the best model proved to be too difficult question for regex - 127 papers identified as not stating best method compared to 20 based on full-text analysis by LLM \ref{tab:detailed-best-model-summary}.

\begin{table}
\centering
\caption{Is there information in full text about:}
\label{tab:yes_no_counts}
\rowcolors{2}{gray!25}{white}  
\begin{tabular}{lrrrrr}
\rowcolor{gray!50}
\toprule
{} &  Model Comparison &  HPO &  Data frequency &  Loss function &  Best model \\
\midrule
\textbf{No } &           8 &                           50 &              39 &             65 &         127 \\
\textbf{Yes} &         138 &                          96 &             107 &             81 &          19 \\
\bottomrule
\end{tabular}
\end{table}

To use LLM capabilities we specifically mentioned we want it to use the context. It would be wiser to find specific keywords for Regex search based on the abstract. Then read the context and decide what is the final answer with elaboration. To confirm it followed our guidance, we asked it to summarise logic afterward \ref{LLM Logic}.

The results prove the efficiency of keyword-based searches, such as regex-based, with domain expert knowledge used to select them, as an efficient way to do filtering and find some simple information. However, more complex questions require attention - a small batch of one paper, which then can be filtered down to elaborations on the most crucial parts. \ref{Modele} shows that such a method with LLM can be an insightful and effective way to use in research.

\subsubsection{Errors and consistency}

The most common error encountered was due to excessively large files, such as loading a book with over 100 pages. In some instances, papers were too mathematical for GPT to parse and understand accurately. Parsing such files did not necessarily produce an error; instead, the model either attempted to list chapters as different papers or began to hallucinate based on its partial understanding, as found in research (e.g. \cite{RNli2023halu}). Following their findings, we provide external domain knowledge and break the task by adding reasoning steps.

Excluding books (7 instances, which were easy to filter out), there were 4 errors out of 150 files. To maintain accuracy, it was crucial to send files in small batches—preferably one by one— as larger batches led to confusion and loss of context in GPT. Following best practices from social experiments, we included questions to check GPT's attention (such as asking it to summarise the task at hand), which it passed.

There are idiosyncratic risks associated with relying on a single LLM. For example, GPT-3.5 Turbo 23.03 frequently identified the need for hyperparameter optimization (HPO) while rarely recognizing model comparison. Consequently, when abstracts did not explicitly mention HPO but the models required proper tuning, ChatGPT might incorrectly affirm that HPO was performed. Conversely, GPT-4.0 demonstrated more strict criteria for HPO (33 fewer papers identified) but recognized more instances of model comparison (46 more papers). In the full-text analysis, GPT-3.5 performed better on HPO-related questions, while GPT-4.0 excelled in identifying model comparisons. This domain-specific comparison of performance over time, which shows irregularities, is in line with findings in \cite{tu2024chatlog}.

Another issue observed was the lack of consistency in the LLM's responses. For example, it sometimes did not consider arbitrarily selected benchmarks (e.g., buy and hold) as model comparisons, treating them merely as sanity checks, whereas in other abstracts it did. Occasionally, it treated parameter tuning as HPO. However, analyzing longer answers revealed that the LLM's certainty varied. To address this, we incorporated the elaborations provided by the LLM in our full-text analysis, which enhanced the reliability of the responses.


\section{Conclusions}\label{konki}

In summary, our work presents a replicable methodology that combines state-of-the-art NLP techniques and LLMs to perform large-scale, automated literature reviews. By focusing on algorithmic trading as a case study, we demonstrated the effectiveness of this approach in uncovering trends, patterns, and insights that contribute to a deeper understanding of the field. The innovative use of advanced methodologies not only streamlines the review process but also opens new avenues for research by providing a more granular and dynamic analysis of scientific literature.

\subsection{Case. Algorithmic trading literature review}
Recent advancements in computer science and natural language processing have enabled researchers to access vast databases of scientific papers and narrow them down to their areas of interest. In our study of algorithmic investing strategies, we used a keyword-based approach to filter a large dataset of research papers. Our analysis revealed that algorithmic trading has become increasingly popular over time, particularly between 1990 and 2010. In recent years, shorter time horizons have gained popularity, driven by cheaper computational power and easier access to relevant data. While stocks and indices are the most commonly studied assets, other asset classes have experienced spikes in popularity during certain periods, such as the oil crisis of 2014-2016 or the rise of cryptocurrencies after 2018.

Machine learning-based techniques have become the most widely tested statistical models in the field. Our topic modeling analysis revealed major trends in contemporary research and identified the topic of comparing various algorithms and models, particularly those based on machine learning. While keyword-based approaches are useful for finding popular methods and their intersections, they have limitations in answering questions about which models outperform others in general. 

Full-text analysis has confirmed that HPO, while not often being the main focus of the study, is performed in the vast majority of the papers. Recent studies (\cite{RN205, RN206}) have shown that many fine-tuned algorithms are sensitive to changes in hyperparameters, so it is important to be cautious about the robustness of some methods.

Deep learning models prove to be the most promising models for algorithmic trading, closely followed by traditional statistical models. However, there are many successful neural networks, especially recurrent ones, and there is plenty of research applying reinforcement learning or ensemble methods. 

\subsection{LLM for literature review}

We demonstrated how an automated approach, enhanced by NLP and LLMs, can effectively assist in literature review. By filtering one of the largest databases of research publications we were able to analyze vast amounts of data that would be infeasible to process manually. 

Both regex-based filtering and LLM proved to be successful and useful in refining a huge corpus of research papers. Abstracts, while giving some insight into the study, often omit parts that are found later in the paper (e.g., details about models, HPO, or data). Furthermore, the results varied from version to version, showing inconsistencies reported in previous studies.

We showed how LLMs can enhance the quantitative literature review process by understanding complex concepts and providing nuanced analyses that account for factors like model comparisons and dataset variations. However, we noticed that ChatGPT without reasoning steps tends to oversimplify the problem. By breaking the task - reading and understanding research papers - into simpler parts and guiding the process, we were able to extract nuanced knowledge about used models, datasets, or (loss) functions. This work proved that full paper analysis with LLM can be a sophisticated method of knowledge extraction.

The study confirmed the added value of a step-by-step approach. The most accurate results were achieved by grouping the papers into small batches, we were able to first extract the information in a particular context and save it as elaboration, which was then used in further steps of the analysis.

\section{Declarations}
Large Language Models, namely ChatGPT 3.5 and 4o, were used in this research for evaluation in \ref{konki}, as well as for text, code, and table polishing.

No funding was received to assist with the preparation of this manuscript. The authors have no competing interests to declare that are relevant to the content of this article.

\begin{appendices}
\section{Appendix}
\subsection{Classification and Regex for models}
\label{sub:reg}
Here is the list of regular expressions for each topic: 

\textbf{Linear models}: ordinary least square$\vert$ OLS$\vert$ linear model.$\vert$ lasso$\vert$ ridge

\textbf{Machine Learning}: random forest$\vert$ decision tree.$\vert$ regression tree.$\vert$ xgboost$\vert$ boosting$\vert$ extreme gradient$\vert$ LSTM$\vert$ Long.short.term.$\vert$ support vector regressions$\vert$ SVR$\vert$ support vector machine$\vert$ SVM$\vert$ k.nearest neighbour.$\vert$ knn$\vert$ clustering algo.$\vert$ mapping algo.$\vert$ neural network$\vert$ (imitation$\vert$ reinforcement$\vert$ unsupervised) learning

\textbf{Time series}: GLM$\vert$ Generalized linear model$\vert$ (Poisson(.?point)?$\vert$ Gaussian$\vert$ Normal) (proces.$\vert$ regress.)$\vert$ (s)?ar(i)?ma(x)?$\vert$ garch

\subsection{Asset classes and venues in topic Neural Network Trading}
\begin{table}
\begin{center}

    \caption{Asset classes and venues in topic 16}
    \label{tab:what}

\begin{minipage}[c]{0.21\textwidth}
\rowcolors{2}{gray!25}{white}  
\begin{tabular}{|c|c|}
\hline
\rowcolor{gray!50}
\textbf{Asset} & \textbf{Count} \\ \hline
\text{Stocks} & \text{104} \\ \hline
\text{Indices} & \text{61} \\ \hline
\text{Commodities} & \text{14} \\ \hline
\text{Currencies} & \text{11} \\ \hline
\text{Bonds} & \text{3} \\ \hline
\text{Cryptos} & \text{3} \\ \hline
\end{tabular}

\end{minipage}
\begin{minipage}[c]{0.25\textwidth}
\centering
\rowcolors{2}{gray!25}{white}  
\begin{tabular}{|c|c|}
\hline
\rowcolor{gray!50}
\textbf{Market Indices/Location} & \textbf{Count} \\ [0.5ex]
\hline
S\&P & 16 \\ \hline
EU & 8 \\
\hline
Hang Seng \& other Chinese & 8 \\ [0.25ex]
\hline
KOSPI/Korea & 7 \\
\hline
NYSE & 5 \\
\hline
DJIA & 5 \\
\hline
Bovespa/Brazilian & 3 \\
\hline
Nikkei/Japan & 3 \\
\hline
other US & 2 \\ 
\hline
\end{tabular}
\end{minipage}
\end{center}
\end{table}

While we were able to find the traded asset class, the identification of the venue based on keyword search failed to deliver meaningful results based on abstracts. By human validation, we confirmed that in over two-thirds of this reduced dataset, there is no mention of the particular assets. 

\begin{table}
\centering

    \caption{Techniques used in topic 16: Neural Network Trading}
    \label{tab:how}
\rowcolors{2}{gray!25}{white}  
\begin{tabular}{|l|r|}
\hline
\rowcolor{gray!50}
\textbf{Model} & \textbf{Count} \\ \hline
\rowcolor{gray!25}Neural Network (NN) & 80 \\
\rowcolor{white}Imitation Learning, Reinforcement Learning, Q-Learning/Network, & \\
Actor-Critique, A3C & 62 \\
Machine Learning & 48 \\
Data Mining, Rough Set, Fuzzy & 37 \\
Technical Analysis (TA), Technical Indicator, 
MACD, Oscillator & 36 \\
K-Nearest Neighbour (KNN), Clustering Algo, 
Mapping Algo, Pattern & 27 \\
Rule-based system & 26 \\
Support Vector Regression (SVR), 
Support Vector Machine (SVM) & 25 \\
GLM, Classification, 
Logistic, Multinomial Regression   & 22 \\
Buy and Sell, Buy Sell, Buy and Hold (B\&H) & 18 \\
Random Forest, Decision Tree,
Regression Tree, CART, CHAID & 16 \\
Long Short-Term Memory (LSTM) & 13 \\
Ordinary least square (OLS), linear model & 11 \\
Ensemble, voting & 9 \\
Convolutional Neural Network (CNN) & 8 \\
Fourier and Kernel tricks & 8 \\
Recurrent Neural Network (RNN) & 7 \\
XGBoost, Boosting, Extreme Gradient & 6 \\
Genetic algorithms & 6 \\
Correspondence Analysis (CA) & 5 \\
Principal Component Analysis (PCA),  
encoder, autoencoder & 5 \\
SARIMA, GARCH & 5 \\
Extreme Learning Machine (ELM) & 4 \\
Stochastic Gradient Descent (SGD) & 2 \\
Hyperparameter, 
Hyperparameter Optimization (HPO)  & 2 \\
Lasso, Ridge & 2 \\
Natural Language Processing (NLP) & 2 \\
eXplainable AI & 1 \\
Residual Neural Network (ResNet) & 0 \\
\hline
\end{tabular}
\end{table}

\subsection{Topics per year}
\begin{figure}
\centerline{\includegraphics[angle=270, width=0.7\textwidth]{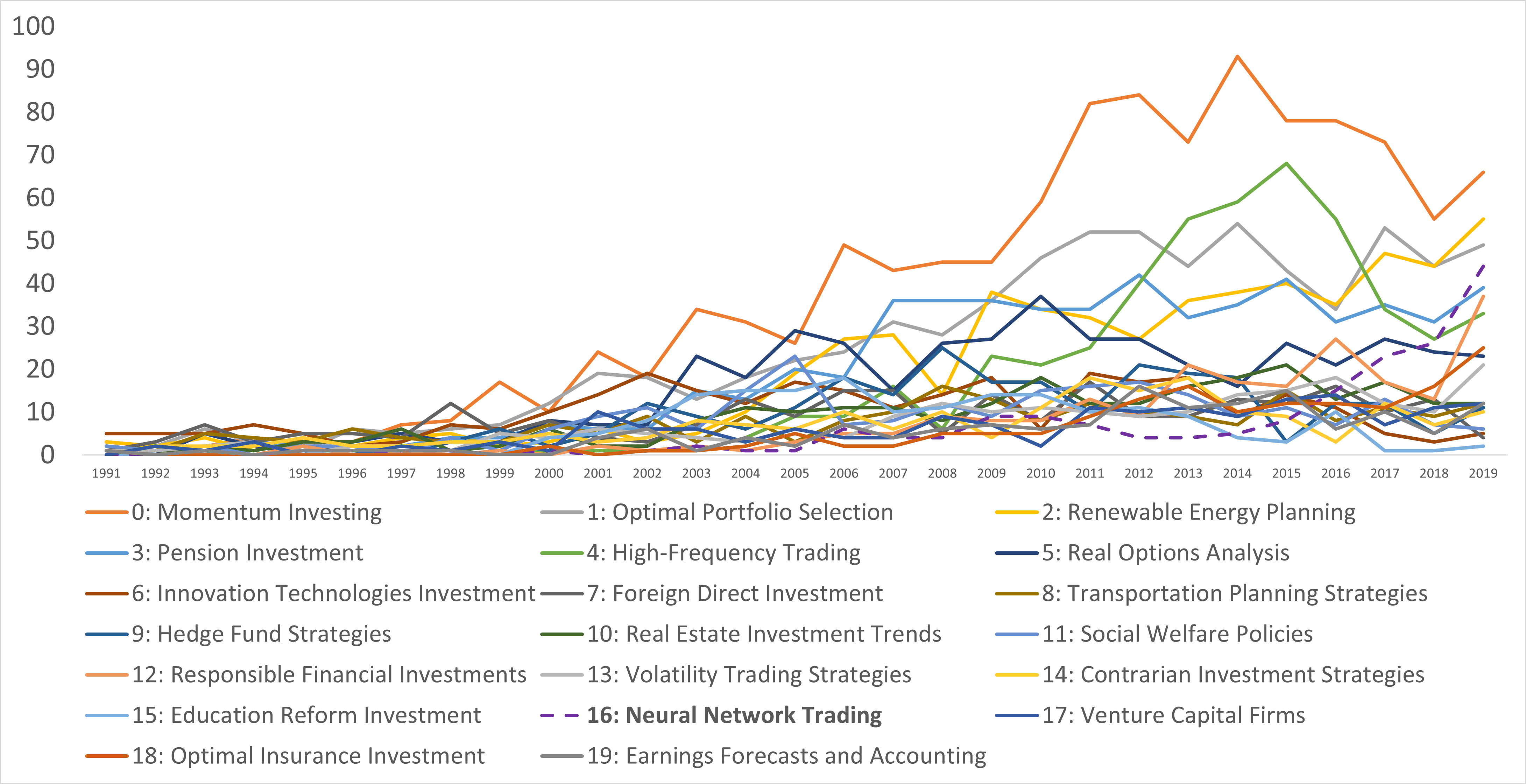}}
\caption{\label{fig:Topics per year}The trends of 20 main topics}
\end{figure}
\clearpage

\subsection{Lazy GPT regex}\label{lazyGPT}
\lstset{frame=tb,
  language=Python,
  breaklines=true,
  postbreak=\mbox{\textcolor{red}{$\hookrightarrow$}\space}}
\begin{lstlisting}
import re
import pdfplumber

def extract_text_from_pdf(pdf_path):
    with pdfplumber.open(pdf_path) as pdf:
        text = ""
        for page in pdf.pages:
            text += page.extract_text()
    return text

def analyze_text(text):
    results = {
        "comparison": {"yes_no": "No", "elaboration": ""},
        "hyperparameter_optimization": {"yes_no": "No", "elaboration": ""},
        "data_frequency": {"yes_no": "No", "elaboration": ""},
        "loss_function": {"yes_no": "No", "elaboration": ""},
        "best_model": {"yes_no": "No", "elaboration": ""}
    }

    # Comparison of Different Models or Methods
    if re.search(r'comparison|compare|benchmark|evaluate|versus|comparison study|side-by-side|comparative analysis', text, re.IGNORECASE):
        results["comparison"]["yes_no"] = "Yes"
        results["comparison"]["elaboration"] = extract_comparison_details(text)
    
    # Hyperparameter Optimization
    if re.search(r'hyperparameter|tuning|optimization|grid search|random search|bayesian optimization|hyperparameter tuning|parameter search|hyper-optimization', text, re.IGNORECASE):
        results["hyperparameter_optimization"]["yes_no"] = "Yes"
        results["hyperparameter_optimization"]["elaboration"] = extract_hyperparameter_details(text)
    
    # Frequency of Data Used
    frequency_match = re.search(r'daily|weekly|monthly|minute-level|hourly|annually|yearly|bi-weekly|quarterly', text, re.IGNORECASE)
    if frequency_match:
        results["data_frequency"]["yes_no"] = "Yes"
        results["data_frequency"]["elaboration"] = frequency_match.group(0)
    
    # Loss Function
    loss_function_match = re.search(r'mean squared error|mse|mean absolute error|mae|cross-entropy|log loss|hinge loss|squared loss|absolute error|mean bias', text, re.IGNORECASE)
    if loss_function_match:
        results["loss_function"]["yes_no"] = "Yes"
        results["loss_function"]["elaboration"] = loss_function_match.group(0)
    
    # Best Model
    best_model_match = re.search(r'best model|optimal model|most accurate|highest performing|top model|leading model|best-performing', text, re.IGNORECASE)
    if best_model_match:
        results["best_model"]["yes_no"] = "Yes"
        results["best_model"]["elaboration"] = extract_best_model(text)
    
    return results

def extract_comparison_details(text):
    # Implement a detailed extraction logic for comparison details
    comparison_sentences = re.findall(r'comparison|compare|benchmark|evaluate|versus|comparison study|side-by-side|comparative analysis.*?\.', text, re.IGNORECASE)
    return " ".join(comparison_sentences)

def extract_hyperparameter_details(text):
    # Implement a detailed extraction logic for hyperparameter details
    hyperparameter_sentences = re.findall(r'hyperparameter|tuning|optimization|grid search|random search|bayesian optimization|hyperparameter tuning|parameter search|hyper-optimization.*?\.', text, re.IGNORECASE)
    return " ".join(hyperparameter_sentences)

def extract_best_model(text):
    # Implement a detailed extraction logic for best model
    best_model_sentences = re.findall(r'best model|optimal model|most accurate|highest performing|top model|leading model|best-performing.*?\.', text, re.IGNORECASE)
    return " ".join(best_model_sentences)

# Example usage
pdf_path = "example_research_paper.pdf"
text = extract_text_from_pdf(pdf_path)
results = analyze_text(text)
print(results)
\end{lstlisting}

\subsection{LLM Logic}\label{LLM Logic}
\lstset{frame=b,
language={},
  breaklines=true,
   postbreak=\space}
\begin{lstlisting}
Reviewing the Abstract: I examined the abstract of the paper to get an overview of its focus, methods, and key findings.

Identifying Comparisons: I looked for any mention of comparisons between different models or methods within the abstract and any additional text available from the paper. This included looking for keywords such as "compare," "comparison," "evaluate," "versus," and "against."

Hyperparameter Optimization: I searched for information on any tuning or optimization of hyperparameters. This typically involves looking for terms like "optimize," "hyperparameter," "parameter tuning," and "settings."

Frequency of Data: I checked for any mention of the data frequency used in the study. This could include daily, weekly, monthly, or any other specific time intervals mentioned in relation to the data.

Loss Function: I looked for any explicit mention of a loss function used in the paper. If a specific loss function was not mentioned, I inferred the optimization criteria from the context, such as the focus on maximizing returns or minimizing errors.

Best Model: I reviewed the findings to identify which model or configuration was reported as the best performing one. This often involves looking for terms like "best," "optimal," "highest performance," and specific model names or configurations.
\end{lstlisting}

\subsection{Unique time horizons found in Neural Network Trading topic}\label{App:Czasy}
\begin{itemize}
\item Intraday
 \begin{itemize}
    \item    'Minute', 'Milliseconds', 'Intra-day', 'High-frequency (minute-level)', 'Hourly', 
        '30-minute', 'High-frequency', 'Intraday', '5-minute intervals', '15-minute', 
        'High Frequency', '30-minute intervals', '5, 10, and 15-minute intervals', 
        '10-minute intervals', 'Tick-level (every microsecond)', '1-minute intervals', 
        'Tick-level (microseconds)', 'High-frequency financial data sampled at an interval of one minute', 
        'High-frequency financial data sampled at one-minute intervals', 'Minute-level', 
        'High-frequency (5-minute intervals)'
\end{itemize}
\item Daily
 \begin{itemize}
 \item 'Daily', 'Daily and weekly', 'Daily and minute-level', 'Daily and 15-minute intervals', 
        'Daily and Monthly', 'Daily, Monthly, Yearly'
 \end{itemize}
\item Longer 
 \begin{itemize}
\item 'Yearly', 'Quarterly', 'Monthly', 'Weekly', 'Various'
\end{itemize}
\end{itemize}

\subsection{Grouping of loss function found in Neural Network Trading topic}\label{Lossy}
\begin{enumerate}
    \item \textbf{Mean Squared Error (MSE) Related:}
    \begin{itemize}
        \item Mean Squared Error (MSE)
        \item Mean Square Error (MSE)
        \item Mean Squared Error with penalizing coefficient
        \item Sum of Square Errors
        \item Mean Squared Forecast Error (MSFE)
        \item Mean Squared Error (MSE) and Cross-Entropy Loss
    \end{itemize}

    \item \textbf{Root Mean Squared Error (RMSE) Related:}
    \begin{itemize}
        \item Root Mean Square Error (RMSE)
        \item RMSE
        \item RMSE and MAPE
        \item RMSE, MAE, MAPE, Theil's U (U1, U2)
    \end{itemize}

    \item \textbf{Cross-Entropy Related:}
    \begin{itemize}
        \item Cross-entropy loss
        \item Cross-Entropy
        \item Binary Cross-Entropy
        \item Categorical Crossentropy
        \item Cross-entropy
        \item Cross-Entropy Loss
        \item Softmax loss function
    \end{itemize}

    \item \textbf{Mean Absolute Percentage Error (MAPE) Related:}
    \begin{itemize}
        \item MAPE
        \item Mean Absolute Percentage Error (MAPE), Directional Accuracy (DA), Theil's U, Average Relative Variance (ARV)
    \end{itemize}

    \item \textbf{Sharpe Ratio Related:}
    \begin{itemize}
        \item Sharpe Ratio
        \item Differential Sharpe Ratio
        \item Sharpe Ratio and Mean Squared Drawdown (MSDD)
        \item Sharpe Ratio Maximization
    \end{itemize}

    \item \textbf{Other Common Loss Functions:}
    \begin{itemize}
        \item Accuracy
        \item Classification Error
        \item Negative Log-Likelihood
        \item Percentage Error
        \item $\epsilon$-insensitive Loss Function ($\epsilon$-ILF)
    \end{itemize}

    \item \textbf{Specialized/Custom Loss Functions:}
    \begin{itemize}
        \item Cost function with a regularization term
        \item Arctangent Cost Function
        \item Structural loss
        \item Reward function and temporal difference error for DDQN, clipped objective function for PPO
        \item Combination of loss functions for actor and critic networks
        \item Optimization criterion based on annualized rate of return, annualized standard deviation, and maximum drawdown
        \item Minimization of the smallest singular vector
        \item Profitability metrics (e.g., return on investment)
        \item Quadratic criterion
        \item Return on Investment (ROI)
        \item Wasserstein distance with Gradient Penalty
    \end{itemize}

\end{enumerate}

\subsection{The best models with explanation and grouped}\label{Modele}
\begin{enumerate}
    \item \textbf{Deep Learning Models}
    \begin{itemize}
        \item Integrated CNN and Deep Learning: Integrated CNN with higher prediction accuracy and cumulative yield.
        \item DeepLearninH2gO Agent: Outperforms MLP and B\&H.
        \item Deep Neural Network (DNN):
        \begin{itemize}
            \item Best performing with 5 hidden layers and sliding window size of 3 minutes.
            \item Uses a neural network ensemble to predict stock returns.
            \item Found to be the best model for predicting financial market movement directions.
            \item Using Stacked Denoising Autoencoders (SdAE) outperformed other models.
            \item Outperforms traditional models with an out-of-sample Sharpe ratio of 2.6.
            \item With a small window size showed the highest directional accuracy and profitability.
            \item Showed superior performance in predicting the direction of financial market movements, achieving up to 68\% accuracy.
        \end{itemize}
        \item TABL (Temporal Attention-Augmented Bilinear Network): Outperforms CNN and LSTM.
        \item CNN with GAF Mapping: Achieved the highest accuracy.
        \item CNN-TA: Outperformed Buy \& Hold, RSI, SMA, LSTM, and MLP regression models.
        \item VLSTM: Outperforms vanilla LSTM, LSTM with attention, multi-scale LSTM, and MLP in terms of mean F1 score.
        \item DLNN: Outperformed the ZIP trading algorithm in live trading tests.
        \item Deep Learning Model:
        \begin{itemize}
            \item Outperforms traditional machine learning methods such as ARIMA and SVM.
            \item Particularly Recurrent Neural Networks (RNNs) and Long Short-Term Memory (LSTM) networks outperformed traditional models.
        \end{itemize}
        \item VAE-LSTM: Demonstrated the highest prediction accuracy and rate of return.
        \item Autoencoder Network: Performed best in terms of profitability in trading simulations.
        \item AANN with Bagging Approach: Showed better performance in terms of trading profits and detection accuracy compared to a buy-and-hold strategy.
        \item Dual Deep Learning Agents: Showed superior performance in option pricing and bid-ask spread estimation.
        \item UFCNN: Showed superior performance in time-series modeling tasks.
        \item CNN-LSTM: Showed superior performance in terms of an annualized rate of return and maximum retracement.
        \item Deep Reinforcement Learning with Sentiment Analysis: Showed superior performance in profitability and risk-adjusted returns.
    \end{itemize}

    \item \textbf{Neural Networks (NN)}
    \begin{itemize}
        \item General Neural Network: Superior returns compared to traditional value strategies.
        \item Neural Network with Specific Topologies:
        \begin{itemize}
            \item Three hidden layers (4, 8, 4 nodes) with inverse tangent and sigmoid transfer functions.
            \item ANN with topology [6, 10, 1]: 6 and 10 neurons in the first two hidden layers, 1 in output.
            \item Neural Network (15 hidden neurons): Inputs include stochastic oscillator \%K \%D, MACD, RSI index, and backward regressions for 5 and 10 days.
        \end{itemize}
        \item MACD Crossover Neural Network: Outperformed TIPP and ANN models in various scenarios.
        \item Feedforward Neural Network (FFNN): Chosen for universal representation capabilities and fast prediction.
        \item Back-Propagation Neural Network (BPNN):
        \begin{itemize}
            \item Outperforms the Genetic Programming model.
            \item Showed superior performance in terms of forecasting accuracy and profitability for inter-commodity spread trading.
        \end{itemize}
        \item BPN2 (Backpropagation Neural Network with architecture 5-3-3-1): Best performance compared to other models (BPN1, BPN3, MR).
        \item Artificial Neural Network (ANN):
        \begin{itemize}
            \item Based on fundamental analysis (FA) concepts for higher prediction accuracy.
            \item Outperformed the ARMA model in some cases.
        \end{itemize}
        \item Self-evolving Trading Strategy Based on BP Neural Network: Outperforms classical strategies in terms of yield and risk management.
        \item Neural Network-Based Framework:
        \begin{itemize}
            \item Demonstrated better performance in predicting profitable trading actions.
            \item Outperforms traditional moving averages and other statistical measures.
        \end{itemize}
        \item FNN with Reduced Complexity Encoding: Showed the best performance in terms of profitability and trading efficiency.
        \item ANFIS: Showed superior performance in terms of Profit Factor, ROI, Sharpe Ratio, and Sortino Ratio.
        \item ANFIS with Active Investment Strategy: Shows the best performance in stock price prediction.
        \item ANFIS-RL: Showed superior performance in terms of predictive accuracy and trading profitability.
        \item AANN: Showed superior performance in detecting trends and generating profitable trading signals.
        \item Red Ward Neural Network: Superior performance in predicting weekly profitability.
    \end{itemize}

    \item \textbf{Reinforcement Learning (RL)}
    \begin{itemize}
        \item Lipschitz Extension-based RL: Highlighted for performance compared to neural networks.
        \item Deep Q-learning Networks (DQN):
        \begin{itemize}
            \item Outperformed classical time-series momentum strategies.
            \item Outperforms other models in terms of profitability and stability in the stock market investment strategy.
        \end{itemize}
        \item Time-driven Feature-aware Jointly Deep Reinforcement Learning (TFJ-DRL): Outperforms other models in terms of total profits and Sharpe ratio.
        \item Fitted Q Iteration with Extra-Trees Regressor: Outperforms basic Q-learning algorithm in handling continuous state and action spaces.
        \item Dynamic Q EKF with ANN: EKF model with dynamically set Q parameter using ANN outperformed constant Q EKF model.
        \item DRL (DQN and A3C): Found to outperform traditional methods and other machine learning models in terms of risk-adjusted returns.
        \item PPO: Showed better convergence and robustness compared to DDQN.
        \item Asynchronous Advantage Actor-Critic (A3C) Method: Demonstrated a stable winning strategy with high profitability and risk management.
        \item Deep Reinforcement Learning with Sentiment Analysis: Showed superior performance in profitability and risk-adjusted returns.
        \item AlphaStock: Showed superior performance in terms of risk-adjusted returns, adaptability to diverse market states, and control of extreme losses.
    \end{itemize}

    \item \textbf{Traditional Machine Learning Models}
    \begin{itemize}
        \item Distance-based Model: Feature-weighted Euclidean distance to the centroid of a training cluster.
        \item Random Forest:
        \begin{itemize}
            \item Adaptability to non-stationary time series data.
            \item Produced the most accurate forecasts and highest abnormal returns.
            \item Showed the best performance in predicting stock price movements.
            \item Combined with boosting algorithms, showed superior performance in detecting economic turning points.
        \end{itemize}
        \item Decision Trees (ID3 Algorithm): Constructed based on financial indicators.
        \item Logistic Regression: Outperformed buy-and-hold and dual momentum strategies.
        \item Multi-classifier System: kNN, Logistic Regression, Naive Bayes, Decision Tree, SVM with genetic algorithms.
        \item CHAID: Best prediction accuracy of 85.64\%.
        \item Linear Regression (LR): Outperforms Support Vector Regression (SVR) in short-term prediction.
        \item Heuristic Forecasting Model (HFM): Outperformed buy-and-hold strategy and non-heuristic forecasting model.
        \item Manifold Learning: Found to yield promising results for FX forecasting.
        \item Cooperative Learning Model: Group knowledge refinement learning model (combination of XCS and neural network).
        \item Extended Hill Climbing (EHC): Effective with lower computation time compared to Exhaustive Search.
        \item Principal Component Regression (PCR): Highest predictive performance, best hit ratio, and $R^2$.
        \item Kernel Price Pattern Trading (KPPT) System: Utilizes a kernel-based approach to predict price patterns.
        \item Zero-Truncated Poisson Mixture Model (ZTP): Outperforms Poisson and Negative Binomial mixture models.
        \item Incremental SVR: Outperforms batch-mode and individual experts.
        \item Differential Evolution Method (DEM): The best balance between computation time and strategy performance.
        \item Gradient Boosting Decision Tree (GBDT): Combined with multi-view feature construction showed superior performance.
        \item Ensemble of SVR Models: Showed superior performance in terms of risk-adjusted returns and profitability.
    \end{itemize}

    \item \textbf{Support Vector Machine (SVM) Models}
    \begin{itemize}
        \item PCA-WSVM: Outperforms WSVM, PCA-ANN, and BHS.
        \item PLR–FW-WSVM: Outperformed PLR–WSVM and PLR–ANN.
        \item Support Vector Machine (SVM):
        \begin{itemize}
            \item Radial basis function (RBF) kernel outperformed logistic regression.
            \item Integrated with GARCH and VPIN, effective in predicting market liquidity and returns.
            \item With Polynomial Kernel: Showed the highest accuracy and returns.
            \item With Radial Basis Function Kernel: Superior performance in volatility forecasting compared to traditional GARCH models.
            \item Based Strategy: Showed competitive performance but did not outperform the equally weighted portfolio strategy (EqW).
        \end{itemize}
        \item ABC-ANFIS-SVM: The hybrid model showed superior performance in terms of accuracy and quality.
    \end{itemize}

    \item \textbf{Rough Sets}
    \begin{itemize}
        \item Rough Sets with LEM2 Algorithm: Outperforms other methods in accuracy and fewer attributes.
        \item Rough Set Analysis: Used to generate trading rules.
        \item Rough Set-Based Rule Generation: High return rates with trend coordination.
        \item Rough Set-Based Real-time Rule-Based Trading System (RRTS): Chosen as the best model.
        \item Rough Set-Based Rule Extraction: Higher return rates compared to traditional technical analysis methods.
        \item S-Rough Sets: More effective than Z. Pawlak rough sets in dynamic information recognition.
        \item Rough Sets Classifier: Chosen as the best model due to handling vagueness, uncertainty, and incomplete data.
    \end{itemize}

    \item \textbf{Recurrent Neural Networks and extensions}
    \begin{itemize}
        \item NARX Network: Outperforms SVM when combined with ICA.
        \item SFM Network: Outperforms AR and LSTM.
        \item CR (Candlestick-based RRL): Outperforms basic RRL, ZI, and BH models.
        \item VG-RAM WNN: Outperformed ARNN predictors in computational efficiency.
        \item SAF-ARC-MMSGD: Outperforms other models in terms of convergence speed and robustness against impulsive noise.
        \item DeepMTA: Outperforms Logistic Regression, Hidden Markov Model, and Dual-Attention RNN.
        \item LSTM with OSTSC: Improved performance metrics compared to the model without oversampling.
        \item RF-WMGEPSVM: Outperforms other strategies in terms of ROR, MDD, and PP across various market scenarios.
        \item EMD-ELM-PLUS: Outperforms EMD-ELM-ELM and single models in terms of RMSE, MAPE, and DA.
        \item PMTS with DTW: Outperformed other methods in terms of trading profitability and stability.
        \item FA-FFLANN with RLS: Outperforms other models in terms of MAPE, DA, Theil’s U, and ARV.
        \item WGAN-GP: Superior performance in generating realistic financial time series.
        \item PLR-IRF and DRNN-Based Model: Showed higher prediction accuracy and lower critical error rate.
    \end{itemize}

    \item \textbf{Ensemble Models}
    \begin{itemize}
        \item Ensemble Model: Combines predictions of multiple models (SVM, decision trees, neural networks).
        \item Hybrid Model: Combining PNN, rough sets, and C4.5 decision tree.
        \item Random Forest and Boosting Algorithms: Superior performance in detecting economic turning points.
    \end{itemize}

    \item \textbf{Hybrid and Composite Models}
    \begin{itemize}
        \item PLR–FW-WSVM: Outperformed PLR–WSVM and PLR–ANN in accuracy and profit.
        \item Multi-classifier System: kNN, LR, NB, DT, SVM with genetic algorithms.
        \item MACD crossover neural network: Outperformed TIPP and ANN in various scenarios.
        \item DeepMTA: Outperforms Logistic Regression, Hidden Markov Model, and Dual-Attention RNN.
        \item GABPN: Outperforms BPN and multiple regression models.
        \item Hybrid Fuzzy Inference System (HyFIS): Achieved the highest hit ratio and best cumulative wealth performance.
        \item ABC-ANFIS-SVM: The hybrid model showed superior performance in terms of accuracy and quality.
        \item MLP: Showed superior performance in prediction accuracy and profitability.
    \end{itemize}

        \item \textbf{Specialized Models}
     \begin{itemize}
        \item Models with Negative Coefficients and Small Intercepts: Perform well in terms of profitability and hit ratios.
        \item Polynomial Solver: Shows promising results in minimizing prediction error.
        \item eFSM-Based Straddle Trading System: Superior performance in volatility prediction and trading profitability.
        \item Machine Learning-Based Synthetic Data Generation: More effective in addressing issues of small data and outliers.
        \item Machine Learning-Based Trading Algorithms: Outperform traditional models and human traders.
        \item AML-Based Strategy: Superior performance in the probability of correct selection and efficiency.
        \item ELM-SVR Combined with Kalman Filter: Showed the best performance in terms of annualized returns, Sharpe ratio, and reduced volatility.
        \item WiSARD: Improved trading performance with higher win ratios and expectancies.
        \item k-NN with Fuzzy Candlestick Patterns: Showed superior performance in predicting future market behavior.
        \item SOM-Based Strategy: Showed superior performance in profitability and accuracy of trading signals.
    \end{itemize}

    \item \textbf{Others}
        \begin{itemize}
        \item Not applicable: The concept of the best model does not apply.
      \end{itemize}
\end{enumerate}
\clearpage

\subsection{Abbreviations}\label{Abb}
\begin{table}[htbp]
\caption{List of Abbreviations}
\label{tab:abbreviations}
\rowcolors{2}{gray!25}{white}  
\begin{tabular}{|l|l|}
\hline
\rowcolor{gray!50}
\textbf{Abbreviation} & \textbf{Meaning} \\ \hline
BERT & Bidirectional Encoder Representations from Transformers \\ \hline
GPT & Generative Pre-trained Transformer  \\ \hline
LSTM & Long Short-Term Memory \\ \hline
RNN & Recurrent Neural Network \\ \hline
HPO & Hyperparameter Optimization \\ \hline
HDBSCAN & Hierarchical Density-Based Spatial Clustering of Applications with Noise \\ \hline
UMAP & Uniform Manifold Approximation and Projection \\ \hline
PCA & Principal Component Analysis \\ \hline
t-SNE & t-distributed Stochastic Neighbor Embedding \\ \hline
TF-IDF & Term Frequency–Inverse Document Frequency \\ \hline
HFT & High-Frequency Trading \\ \hline
LSTM & Long Short-Term Memory \\ \hline
RQ & Research Question \\ \hline
ML & Machine Learning \\ \hline
TA & Technical Analysis \\ \hline
KNN & K-Nearest Neighbour \\ \hline
SVR & Support Vector Regression \\ \hline
SVM & Support Vector Machine \\ \hline
GLM & Generalized Linear Model \\ \hline
B\&H & Buy and Hold \\ \hline
RF & Random Forest \\ \hline
DT & Decision Tree \\ \hline
CAPM & Capital Asset Pricing Model \\ \hline
CART & Classification and Regression Tree \\ \hline
OLS & Ordinary Least Squares \\ \hline
CNN & Convolutional Neural Network \\ \hline
PCA & Principal Component Analysis \\ \hline
XGBoost & Extreme Gradient Boosting \\ \hline
ELM & Extreme Learning Machine \\ \hline
SGD & Stochastic Gradient Descent \\ \hline
NLP & Natural Language Processing \\ \hline
XAI & eXplainable AI \\ \hline
ResNet & Residual Neural Network \\ \hline
MSE & Mean Squared Error \\ \hline
RMSE & Root Mean Square Error \\ \hline
ROI & Return on Investment \\ \hline
PPO & Proximal Policy Optimization \\ \hline
DDN & Deep Deterministic Network \\ \hline
A3C & Actor-Critic, Asynchronous Advantage Actor-Critic \\ \hline
SARIMA & Seasonal Autoregressive Integrated Moving Average \\ \hline
GARCH & Generalized Autoregressive Conditional Heteroskedasticity \\ \hline
ARIMA & Autoregressive Integrated Moving Average \\ \hline

\end{tabular}
\end{table}

\end{appendices}

\bibliography{ecai.bib}


\begin{thebibliography}{26}
\ifx \bisbn   \undefined \def \bisbn  #1{ISBN #1}\fi
\ifx \binits  \undefined \def \binits#1{#1}\fi
\ifx \bauthor  \undefined \def \bauthor#1{#1}\fi
\ifx \batitle  \undefined \def \batitle#1{#1}\fi
\ifx \bjtitle  \undefined \def \bjtitle#1{#1}\fi
\ifx \bvolume  \undefined \def \bvolume#1{\textbf{#1}}\fi
\ifx \byear  \undefined \def \byear#1{#1}\fi
\ifx \bissue  \undefined \def \bissue#1{#1}\fi
\ifx \bfpage  \undefined \def \bfpage#1{#1}\fi
\ifx \blpage  \undefined \def \blpage #1{#1}\fi
\ifx \burl  \undefined \def \burl#1{\textsf{#1}}\fi
\ifx \doiurl  \undefined \def \doiurl#1{\url{https://doi.org/#1}}\fi
\ifx \betal  \undefined \def \betal{\textit{et al.}}\fi
\ifx \binstitute  \undefined \def \binstitute#1{#1}\fi
\ifx \binstitutionaled  \undefined \def \binstitutionaled#1{#1}\fi
\ifx \bctitle  \undefined \def \bctitle#1{#1}\fi
\ifx \beditor  \undefined \def \beditor#1{#1}\fi
\ifx \bpublisher  \undefined \def \bpublisher#1{#1}\fi
\ifx \bbtitle  \undefined \def \bbtitle#1{#1}\fi
\ifx \bedition  \undefined \def \bedition#1{#1}\fi
\ifx \bseriesno  \undefined \def \bseriesno#1{#1}\fi
\ifx \blocation  \undefined \def \blocation#1{#1}\fi
\ifx \bsertitle  \undefined \def \bsertitle#1{#1}\fi
\ifx \bsnm \undefined \def \bsnm#1{#1}\fi
\ifx \bsuffix \undefined \def \bsuffix#1{#1}\fi
\ifx \bparticle \undefined \def \bparticle#1{#1}\fi
\ifx \barticle \undefined \def \barticle#1{#1}\fi
\bibcommenthead
\ifx \bconfdate \undefined \def \bconfdate #1{#1}\fi
\ifx \botherref \undefined \def \botherref #1{#1}\fi
\ifx \url \undefined \def \url#1{\textsf{#1}}\fi
\ifx \bchapter \undefined \def \bchapter#1{#1}\fi
\ifx \bbook \undefined \def \bbook#1{#1}\fi
\ifx \bcomment \undefined \def \bcomment#1{#1}\fi
\ifx \oauthor \undefined \def \oauthor#1{#1}\fi
\ifx \citeauthoryear \undefined \def \citeauthoryear#1{#1}\fi
\ifx \endbibitem  \undefined \def \endbibitem {}\fi
\ifx \bconflocation  \undefined \def \bconflocation#1{#1}\fi
\ifx \arxivurl  \undefined \def \arxivurl#1{\textsf{#1}}\fi
\csname PreBibitemsHook\endcsname

\bibitem[\protect\citeauthoryear{Bao et~al.}{2019}]{RN195}
\begin{barticle}
\bauthor{\bsnm{Bao}, \binits{Y.}},
\bauthor{\bsnm{Deng}, \binits{Z.}},
\bauthor{\bsnm{Wang}, \binits{Y.}},
\bauthor{\bsnm{Kim}, \binits{H.}},
\bauthor{\bsnm{Armengol}, \binits{V.D.}},
\bauthor{\bsnm{Acevedo}, \binits{F.}},
\bauthor{\bsnm{Ouardaoui}, \binits{N.}},
\bauthor{\bsnm{Wang}, \binits{C.}},
\bauthor{\bsnm{Parmigiani}, \binits{G.}},
\bauthor{\bsnm{Barzilay}, \binits{R.}},
\bauthor{\bsnm{Braun}, \binits{D.}},
\bauthor{\bsnm{Hughes}, \binits{K.S.}}:
\batitle{Using machine learning and natural language processing to review and classify the medical literature on cancer susceptibility genes}.
\bjtitle{JCO Clin Cancer Inform}
\bvolume{3},
\bfpage{1}--\blpage{9}
(\byear{2019})
\doiurl{10.1200/cci.19.00042}
\end{barticle}
\endbibitem

\bibitem[\protect\citeauthoryear{Brown et~al.}{2020}]{RN204}
\begin{barticle}
\bauthor{\bsnm{Brown}, \binits{T.}},
\bauthor{\bsnm{Mann}, \binits{B.}},
\bauthor{\bsnm{Ryder}, \binits{N.}},
\bauthor{\bsnm{Subbiah}, \binits{M.}},
\bauthor{\bsnm{Kaplan}, \binits{J.D.}},
\bauthor{\bsnm{Dhariwal}, \binits{P.}},
\bauthor{\bsnm{Neelakantan}, \binits{A.}},
\bauthor{\bsnm{Shyam}, \binits{P.}},
\bauthor{\bsnm{Sastry}, \binits{G.}},
\bauthor{\bsnm{Askell}, \binits{A.}},
\bauthor{\bsnm{Agarwal}, \binits{S.}},
\bauthor{\bsnm{Herbert-Voss}, \binits{A.}},
\bauthor{\bsnm{Krueger}, \binits{G.}},
\bauthor{\bsnm{Henighan}, \binits{T.}},
\bauthor{\bsnm{Child}, \binits{R.}},
\bauthor{\bsnm{Ramesh}, \binits{A.}},
\bauthor{\bsnm{Ziegler}, \binits{D.}},
\bauthor{\bsnm{Wu}, \binits{J.}},
\bauthor{\bsnm{Winter}, \binits{C.}},
\bauthor{\bsnm{Hesse}, \binits{C.}},
\bauthor{\bsnm{Chen}, \binits{M.}},
\bauthor{\bsnm{Sigler}, \binits{E.}},
\bauthor{\bsnm{Litwin}, \binits{M.}},
\bauthor{\bsnm{Gray}, \binits{S.}},
\bauthor{\bsnm{Chess}, \binits{B.}},
\bauthor{\bsnm{Clark}, \binits{J.}},
\bauthor{\bsnm{Berner}, \binits{C.}},
\bauthor{\bsnm{McCandlish}, \binits{S.}},
\bauthor{\bsnm{Radford}, \binits{A.}},
\bauthor{\bsnm{Sutskever}, \binits{I.}},
\bauthor{\bsnm{Amodei}, \binits{D.}}:
\batitle{{Language Models are Few-Shot Learners}}.
\bjtitle{Advances in Neural Information Processing Systems}
\bvolume{33},
\bfpage{1877}--\blpage{1901}
(\byear{2020})
\doiurl{10.48550/arXiv.2005.14165}
\end{barticle}
\endbibitem

\bibitem[\protect\citeauthoryear{Cachola et~al.}{2020}]{RNCachola}
\begin{botherref}
\oauthor{\bsnm{Cachola}, \binits{I.}},
\oauthor{\bsnm{Lo}, \binits{K.}},
\oauthor{\bsnm{Cohan}, \binits{A.}},
\oauthor{\bsnm{Weld}, \binits{D.S.}}:
{TLDR: Extreme Summarization of Scientific Documents}.
CoRR
\textbf{abs/2004.15011}
(2020)
{\href{https://arxiv.org/abs/2004.15011}{{2004.15011}}}
\end{botherref}
\endbibitem

\bibitem[\protect\citeauthoryear{Devlin et~al.}{2019}]{Bert}
\begin{barticle}
\bauthor{\bsnm{Devlin}, \binits{J.}},
\bauthor{\bsnm{Chang}, \binits{M.-W.}},
\bauthor{\bsnm{Lee}, \binits{K.}},
\bauthor{\bsnm{Toutanova}, \binits{K.}}:
\batitle{{BERT: Pre-training of Deep Bidirectional Transformers for Language Understanding}}.
\bjtitle{arXiv pre-print server}
(\byear{2019})
\doiurl{10.48550/arXiv.1810.04805}
\end{barticle}
\endbibitem

\bibitem[\protect\citeauthoryear{Dowling and Lucey}{2023}]{RN194}
\begin{barticle}
\bauthor{\bsnm{Dowling}, \binits{M.}},
\bauthor{\bsnm{Lucey}, \binits{B.}}:
\batitle{{ChatGPT for (Finance) research: The Bananarama Conjecture}}.
\bjtitle{Finance Research Letters}
\bvolume{53},
\bfpage{103662}
(\byear{2023})
\doiurl{10.1016/j.frl.2023.103662}
\end{barticle}
\endbibitem

\bibitem[\protect\citeauthoryear{Fire and Guestrin}{2019}]{RN165}
\begin{botherref}
\oauthor{\bsnm{Fire}, \binits{M.}},
\oauthor{\bsnm{Guestrin}, \binits{C.}}:
{O}ver-optimization of academic publishing metrics: observing {G}oodhart’s {L}aw in action.
GigaScience
\textbf{8}(6)
(2019)
\doiurl{10.1093/gigascience/giz053}
\end{botherref}
\endbibitem

\bibitem[\protect\citeauthoryear{Ferreira et~al.}{2021}]{RNlitrev}
\begin{barticle}
\bauthor{\bsnm{Ferreira}, \binits{F.G.D.C.}},
\bauthor{\bsnm{Gandomi}, \binits{A.H.}},
\bauthor{\bsnm{Cardoso}, \binits{R.T.N.}}:
\batitle{{Artificial Intelligence Applied to Stock Market Trading: A Review}}.
\bjtitle{IEEE Access}
\bvolume{9},
\bfpage{30898}--\blpage{30917}
(\byear{2021})
\doiurl{10.1109/ACCESS.2021.3058133}
\end{barticle}
\endbibitem

\bibitem[\protect\citeauthoryear{Gunnarsson et~al.}{2024}]{GUNNARSSON2024103221}
\begin{barticle}
\bauthor{\bsnm{Gunnarsson}, \binits{E.S.}},
\bauthor{\bsnm{Isern}, \binits{H.R.}},
\bauthor{\bsnm{Kaloudis}, \binits{A.}},
\bauthor{\bsnm{Risstad}, \binits{M.}},
\bauthor{\bsnm{Vigdel}, \binits{B.}},
\bauthor{\bsnm{Westgaard}, \binits{S.}}:
\batitle{{Prediction of realized volatility and implied volatility indices using AI and machine learning: A review}}.
\bjtitle{International Review of Financial Analysis}
\bvolume{93},
\bfpage{103221}
(\byear{2024})
\doiurl{10.1016/j.irfa.2024.103221}
\end{barticle}
\endbibitem

\bibitem[\protect\citeauthoryear{Grootendorst}{2022}]{RN200}
\begin{barticle}
\bauthor{\bsnm{Grootendorst}, \binits{M.}}:
\batitle{{BERTopic: Neural topic modeling with a class-based TF-IDF procedure}}.
\bjtitle{arXiv preprint arXiv:2203.05794}
(\byear{2022})
\doiurl{10.48550/arXiv.2203.05794}
\end{barticle}
\endbibitem

\bibitem[\protect\citeauthoryear{Garcia et~al.}{2022}]{RN203}
\begin{barticle}
\bauthor{\bsnm{Garcia}, \binits{J.}},
\bauthor{\bsnm{Villavicencio}, \binits{G.}},
\bauthor{\bsnm{Altimiras}, \binits{F.}},
\bauthor{\bsnm{Crawford}, \binits{B.}},
\bauthor{\bsnm{Soto}, \binits{R.}},
\bauthor{\bsnm{Minatogawa}, \binits{V.}},
\bauthor{\bsnm{Franco}, \binits{M.}},
\bauthor{\bsnm{Martínez-Muñoz}, \binits{D.}},
\bauthor{\bsnm{Yepes}, \binits{V.}}:
\batitle{Machine learning techniques applied to construction: {A} hybrid bibliometric analysis of advances and future directions}.
\bjtitle{Automation in Construction}
\bvolume{142},
\bfpage{104532}
(\byear{2022})
\doiurl{10.1016/j.autcon.2022.104532}
\end{barticle}
\endbibitem

\bibitem[\protect\citeauthoryear{Hong et~al.}{2023}]{RN170}
\begin{botherref}
\oauthor{\bsnm{Hong}, \binits{Z.}},
\oauthor{\bsnm{Ajith}, \binits{A.}},
\oauthor{\bsnm{Pauloski}, \binits{J.}},
\oauthor{\bsnm{Duede}, \binits{E.}},
\oauthor{\bsnm{Chard}, \binits{K.}},
\oauthor{\bsnm{Foster}, \binits{I.}}:
{The Diminishing Returns of Masked Language Models to Science}.
Findings of the Association for Computational Linguistics: ACL 2023,
1270--1283
(2023)
\doiurl{10.18653/v1/2023.findings-acl.82}
\end{botherref}
\endbibitem

\bibitem[\protect\citeauthoryear{Joiner et~al.}{2022}]{Joiner}
\begin{botherref}
\oauthor{\bsnm{Joiner}, \binits{D.}},
\oauthor{\bsnm{Vezeau}, \binits{A.}},
\oauthor{\bsnm{Wong}, \binits{A.}},
\oauthor{\bsnm{Hains}, \binits{G.}},
\oauthor{\bsnm{Khmelevsky}, \binits{Y.}}:
{Algorithmic Trading and Short-term Forecast for Financial Time Series with Machine Learning Models; State of the Art and Perspectives}.
2022 IEEE International Conference on Recent Advances in Systems Science and Engineering (RASSE),
1--9
(2022)
\doiurl{10.1109/RASSE54974.2022.9989592}
\end{botherref}
\endbibitem

\bibitem[\protect\citeauthoryear{Li et~al.}{2020}]{RN206}
\begin{barticle}
\bauthor{\bsnm{Li}, \binits{H.}},
\bauthor{\bsnm{Chaudhari}, \binits{P.}},
\bauthor{\bsnm{Yang}, \binits{H.}},
\bauthor{\bsnm{Lam}, \binits{M.}},
\bauthor{\bsnm{Ravichandran}, \binits{A.}},
\bauthor{\bsnm{Bhotika}, \binits{R.}},
\bauthor{\bsnm{Soatto}, \binits{S.}}:
\batitle{{Rethinking the Hyperparameters for Fine-tuning}}.
\bjtitle{arXiv pre-print server}
(\byear{2020})
\doiurl{10.48550/arXiv.2002.11770}
\end{barticle}
\endbibitem

\bibitem[\protect\citeauthoryear{Li et~al.}{2023}]{RNli2023halu}
\begin{botherref}
\oauthor{\bsnm{Li}, \binits{J.}},
\oauthor{\bsnm{Cheng}, \binits{X.}},
\oauthor{\bsnm{Zhao}, \binits{X.}},
\oauthor{\bsnm{Nie}, \binits{J.-Y.}},
\oauthor{\bsnm{Wen}, \binits{J.-R.}}:
{{H}alu{E}val: A Large-Scale Hallucination Evaluation Benchmark for Large Language Models}.
Proceedings of the 2023 Conference on Empirical Methods in Natural Language Processing,
6449--6464
(2023)
\doiurl{10.18653/v1/2023.emnlp-main.397}
\end{botherref}
\endbibitem

\bibitem[\protect\citeauthoryear{Lo et~al.}{2020}]{RN164}
\begin{botherref}
\oauthor{\bsnm{Lo}, \binits{K.}},
\oauthor{\bsnm{Wang}, \binits{L.L.}},
\oauthor{\bsnm{Neumann}, \binits{M.}},
\oauthor{\bsnm{Kinney}, \binits{R.}},
\oauthor{\bsnm{Weld}, \binits{D.}}:
{{S}2{ORC}: The Semantic Scholar Open Research Corpus}.
Proceedings of the 58th Annual Meeting of the Association for Computational Linguistics,
4969--4983
(2020)
\doiurl{10.18653/v1/2020.acl-main.447}
\end{botherref}
\endbibitem

\bibitem[\protect\citeauthoryear{McInnes et~al.}{2017}]{RN199}
\begin{barticle}
\bauthor{\bsnm{McInnes}, \binits{L.}},
\bauthor{\bsnm{Healy}, \binits{J.}},
\bauthor{\bsnm{Astels}, \binits{S.}}:
\batitle{hdbscan: {H}ierarchical density based clustering}.
\bjtitle{The Journal of Open Source Software}
\bvolume{2}(\bissue{11}),
\bfpage{205}
(\byear{2017})
\doiurl{10.21105/joss.00205}
\end{barticle}
\endbibitem

\bibitem[\protect\citeauthoryear{McInnes et~al.}{2018}]{RN198}
\begin{barticle}
\bauthor{\bsnm{McInnes}, \binits{L.}},
\bauthor{\bsnm{Healy}, \binits{J.}},
\bauthor{\bsnm{Saul}, \binits{N.}},
\bauthor{\bsnm{Großberger}, \binits{L.}}:
\batitle{{UMAP: Uniform Manifold Approximation and Projection}}.
\bjtitle{Journal of Open Source Software}
\bvolume{3}(\bissue{29}),
\bfpage{861}
(\byear{2018})
\doiurl{10.21105/joss.00861}
\end{barticle}
\endbibitem

\bibitem[\protect\citeauthoryear{Marshall and Wallace}{2019}]{RN188}
\begin{barticle}
\bauthor{\bsnm{Marshall}, \binits{I.J.}},
\bauthor{\bsnm{Wallace}, \binits{B.C.}}:
\batitle{Toward systematic review automation: a practical guide to using machine learning tools in research synthesis}.
\bjtitle{Systematic Reviews}
\bvolume{8}(\bissue{1}),
\bfpage{163}
(\byear{2019})
\doiurl{10.1186/s13643-019-1074-9}
\end{barticle}
\endbibitem

\bibitem[\protect\citeauthoryear{OpenAI et~al.}{2024}]{openai2024gpt4technicalreport}
\begin{botherref}
\oauthor{\bsnm{OpenAI}},
\oauthor{\bsnm{Achiam}, \binits{J.}},
\oauthor{\bsnm{Adler}, \binits{S.}},
\oauthor{\bsnm{Agarwal}, \binits{S.}},
\oauthor{\bsnm{Ahmad}, \binits{L.}},
\oauthor{\bsnm{Akkaya}, \binits{I.}},
\oauthor{\bsnm{Aleman}, \binits{F.L.}},
\oauthor{\bsnm{Almeida}, \binits{D.}},
\oauthor{\bsnm{Altenschmidt}, \binits{J.}},
\oauthor{\bsnm{Altman}, \binits{S.}},
\oauthor{\bsnm{Anadkat}, \binits{S.}},
\oauthor{\bsnm{Avila}, \binits{R.}},
\oauthor{\bsnm{Babuschkin}, \binits{I.}},
\oauthor{\bsnm{Balaji}, \binits{S.}},
\oauthor{\bsnm{Balcom}, \binits{V.}},
\oauthor{\bsnm{Baltescu}, \binits{P.}},
\oauthor{\bsnm{Bao}, \binits{H.}},
\oauthor{\bsnm{Bavarian}, \binits{M.}},
\oauthor{\bsnm{Belgum}, \binits{J.}},
\oauthor{\bsnm{Bello}, \binits{I.}},
\oauthor{\bsnm{Berdine}, \binits{J.}},
\oauthor{\bsnm{Bernadett-Shapiro}, \binits{G.}},
\oauthor{\bsnm{Berner}, \binits{C.}},
\oauthor{\bsnm{Bogdonoff}, \binits{L.}},
\oauthor{\bsnm{Boiko}, \binits{O.}},
\oauthor{\bsnm{Boyd}, \binits{M.}},
\oauthor{\bsnm{Brakman}, \binits{A.-L.}},
\oauthor{\bsnm{Brockman}, \binits{G.}},
\oauthor{\bsnm{Brooks}, \binits{T.}},
\oauthor{\bsnm{Brundage}, \binits{M.}},
\oauthor{\bsnm{Button}, \binits{K.}},
\oauthor{\bsnm{Cai}, \binits{T.}},
\oauthor{\bsnm{Campbell}, \binits{R.}},
\oauthor{\bsnm{Cann}, \binits{A.}},
\oauthor{\bsnm{Carey}, \binits{B.}},
\oauthor{\bsnm{Carlson}, \binits{C.}},
\oauthor{\bsnm{Carmichael}, \binits{R.}},
\oauthor{\bsnm{Chan}, \binits{B.}},
\oauthor{\bsnm{Chang}, \binits{C.}},
\oauthor{\bsnm{Chantzis}, \binits{F.}},
\oauthor{\bsnm{Chen}, \binits{D.}},
\oauthor{\bsnm{Chen}, \binits{S.}},
\oauthor{\bsnm{Chen}, \binits{R.}},
\oauthor{\bsnm{Chen}, \binits{J.}},
\oauthor{\bsnm{Chen}, \binits{M.}},
\oauthor{\bsnm{Chess}, \binits{B.}},
\oauthor{\bsnm{Cho}, \binits{C.}},
\oauthor{\bsnm{Chu}, \binits{C.}},
\oauthor{\bsnm{Chung}, \binits{H.W.}},
\oauthor{\bsnm{Cummings}, \binits{D.}},
\oauthor{\bsnm{Currier}, \binits{J.}},
\oauthor{\bsnm{Dai}, \binits{Y.}},
\oauthor{\bsnm{Decareaux}, \binits{C.}},
\oauthor{\bsnm{Degry}, \binits{T.}},
\oauthor{\bsnm{Deutsch}, \binits{N.}},
\oauthor{\bsnm{Deville}, \binits{D.}},
\oauthor{\bsnm{Dhar}, \binits{A.}},
\oauthor{\bsnm{Dohan}, \binits{D.}},
\oauthor{\bsnm{Dowling}, \binits{S.}},
\oauthor{\bsnm{Dunning}, \binits{S.}},
\oauthor{\bsnm{Ecoffet}, \binits{A.}},
\oauthor{\bsnm{Eleti}, \binits{A.}},
\oauthor{\bsnm{Eloundou}, \binits{T.}},
\oauthor{\bsnm{Farhi}, \binits{D.}},
\oauthor{\bsnm{Fedus}, \binits{L.}},
\oauthor{\bsnm{Felix}, \binits{N.}},
\oauthor{\bsnm{Fishman}, \binits{S.P.}},
\oauthor{\bsnm{Forte}, \binits{J.}},
\oauthor{\bsnm{Fulford}, \binits{I.}},
\oauthor{\bsnm{Gao}, \binits{L.}},
\oauthor{\bsnm{Georges}, \binits{E.}},
\oauthor{\bsnm{Gibson}, \binits{C.}},
\oauthor{\bsnm{Goel}, \binits{V.}},
\oauthor{\bsnm{Gogineni}, \binits{T.}},
\oauthor{\bsnm{Goh}, \binits{G.}},
\oauthor{\bsnm{Gontijo-Lopes}, \binits{R.}},
\oauthor{\bsnm{Gordon}, \binits{J.}},
\oauthor{\bsnm{Grafstein}, \binits{M.}},
\oauthor{\bsnm{Gray}, \binits{S.}},
\oauthor{\bsnm{Greene}, \binits{R.}},
\oauthor{\bsnm{Gross}, \binits{J.}},
\oauthor{\bsnm{Gu}, \binits{S.S.}},
\oauthor{\bsnm{Guo}, \binits{Y.}},
\oauthor{\bsnm{Hallacy}, \binits{C.}},
\oauthor{\bsnm{Han}, \binits{J.}},
\oauthor{\bsnm{Harris}, \binits{J.}},
\oauthor{\bsnm{He}, \binits{Y.}},
\oauthor{\bsnm{Heaton}, \binits{M.}},
\oauthor{\bsnm{Heidecke}, \binits{J.}},
\oauthor{\bsnm{Hesse}, \binits{C.}},
\oauthor{\bsnm{Hickey}, \binits{A.}},
\oauthor{\bsnm{Hickey}, \binits{W.}},
\oauthor{\bsnm{Hoeschele}, \binits{P.}},
\oauthor{\bsnm{Houghton}, \binits{B.}},
\oauthor{\bsnm{Hsu}, \binits{K.}},
\oauthor{\bsnm{Hu}, \binits{S.}},
\oauthor{\bsnm{Hu}, \binits{X.}},
\oauthor{\bsnm{Huizinga}, \binits{J.}},
\oauthor{\bsnm{Jain}, \binits{S.}},
\oauthor{\bsnm{Jain}, \binits{S.}},
\oauthor{\bsnm{Jang}, \binits{J.}},
\oauthor{\bsnm{Jiang}, \binits{A.}},
\oauthor{\bsnm{Jiang}, \binits{R.}},
\oauthor{\bsnm{Jin}, \binits{H.}},
\oauthor{\bsnm{Jin}, \binits{D.}},
\oauthor{\bsnm{Jomoto}, \binits{S.}},
\oauthor{\bsnm{Jonn}, \binits{B.}},
\oauthor{\bsnm{Jun}, \binits{H.}},
\oauthor{\bsnm{Kaftan}, \binits{T.}},
\oauthor{\bsnm{Kaiser}},
\oauthor{\bsnm{Kamali}, \binits{A.}},
\oauthor{\bsnm{Kanitscheider}, \binits{I.}},
\oauthor{\bsnm{Keskar}, \binits{N.S.}},
\oauthor{\bsnm{Khan}, \binits{T.}},
\oauthor{\bsnm{Kilpatrick}, \binits{L.}},
\oauthor{\bsnm{Kim}, \binits{J.W.}},
\oauthor{\bsnm{Kim}, \binits{C.}},
\oauthor{\bsnm{Kim}, \binits{Y.}},
\oauthor{\bsnm{Kirchner}, \binits{J.H.}},
\oauthor{\bsnm{Kiros}, \binits{J.}},
\oauthor{\bsnm{Knight}, \binits{M.}},
\oauthor{\bsnm{Kokotajlo}, \binits{D.}},
\oauthor{\bsnm{Kondraciuk}},
\oauthor{\bsnm{Kondrich}, \binits{A.}},
\oauthor{\bsnm{Konstantinidis}, \binits{A.}},
\oauthor{\bsnm{Kosic}, \binits{K.}},
\oauthor{\bsnm{Krueger}, \binits{G.}},
\oauthor{\bsnm{Kuo}, \binits{V.}},
\oauthor{\bsnm{Lampe}, \binits{M.}},
\oauthor{\bsnm{Lan}, \binits{I.}},
\oauthor{\bsnm{Lee}, \binits{T.}},
\oauthor{\bsnm{Leike}, \binits{J.}},
\oauthor{\bsnm{Leung}, \binits{J.}},
\oauthor{\bsnm{Levy}, \binits{D.}},
\oauthor{\bsnm{Li}, \binits{C.M.}},
\oauthor{\bsnm{Lim}, \binits{R.}},
\oauthor{\bsnm{Lin}, \binits{M.}},
\oauthor{\bsnm{Lin}, \binits{S.}},
\oauthor{\bsnm{Litwin}, \binits{M.}},
\oauthor{\bsnm{Lopez}, \binits{T.}},
\oauthor{\bsnm{Lowe}, \binits{R.}},
\oauthor{\bsnm{Lue}, \binits{P.}},
\oauthor{\bsnm{Makanju}, \binits{A.}},
\oauthor{\bsnm{Malfacini}, \binits{K.}},
\oauthor{\bsnm{Manning}, \binits{S.}},
\oauthor{\bsnm{Markov}, \binits{T.}},
\oauthor{\bsnm{Markovski}, \binits{Y.}},
\oauthor{\bsnm{Martin}, \binits{B.}},
\oauthor{\bsnm{Mayer}, \binits{K.}},
\oauthor{\bsnm{Mayne}, \binits{A.}},
\oauthor{\bsnm{McGrew}, \binits{B.}},
\oauthor{\bsnm{McKinney}, \binits{S.M.}},
\oauthor{\bsnm{McLeavey}, \binits{C.}},
\oauthor{\bsnm{McMillan}, \binits{P.}},
\oauthor{\bsnm{McNeil}, \binits{J.}},
\oauthor{\bsnm{Medina}, \binits{D.}},
\oauthor{\bsnm{Mehta}, \binits{A.}},
\oauthor{\bsnm{Menick}, \binits{J.}},
\oauthor{\bsnm{Metz}, \binits{L.}},
\oauthor{\bsnm{Mishchenko}, \binits{A.}},
\oauthor{\bsnm{Mishkin}, \binits{P.}},
\oauthor{\bsnm{Monaco}, \binits{V.}},
\oauthor{\bsnm{Morikawa}, \binits{E.}},
\oauthor{\bsnm{Mossing}, \binits{D.}},
\oauthor{\bsnm{Mu}, \binits{T.}},
\oauthor{\bsnm{Murati}, \binits{M.}},
\oauthor{\bsnm{Murk}, \binits{O.}},
\oauthor{\bsnm{Mély}, \binits{D.}},
\oauthor{\bsnm{Nair}, \binits{A.}},
\oauthor{\bsnm{Nakano}, \binits{R.}},
\oauthor{\bsnm{Nayak}, \binits{R.}},
\oauthor{\bsnm{Neelakantan}, \binits{A.}},
\oauthor{\bsnm{Ngo}, \binits{R.}},
\oauthor{\bsnm{Noh}, \binits{H.}},
\oauthor{\bsnm{Ouyang}, \binits{L.}},
\oauthor{\bsnm{O'Keefe}, \binits{C.}},
\oauthor{\bsnm{Pachocki}, \binits{J.}},
\oauthor{\bsnm{Paino}, \binits{A.}},
\oauthor{\bsnm{Palermo}, \binits{J.}},
\oauthor{\bsnm{Pantuliano}, \binits{A.}},
\oauthor{\bsnm{Parascandolo}, \binits{G.}},
\oauthor{\bsnm{Parish}, \binits{J.}},
\oauthor{\bsnm{Parparita}, \binits{E.}},
\oauthor{\bsnm{Passos}, \binits{A.}},
\oauthor{\bsnm{Pavlov}, \binits{M.}},
\oauthor{\bsnm{Peng}, \binits{A.}},
\oauthor{\bsnm{Perelman}, \binits{A.}},
\oauthor{\bsnm{Avila Belbute~Peres}, \binits{F.}},
\oauthor{\bsnm{Petrov}, \binits{M.}},
\oauthor{\bsnm{Oliveira~Pinto}, \binits{H.P.}},
\oauthor{\bsnm{Michael}},
\oauthor{\bsnm{Pokorny}},
\oauthor{\bsnm{Pokrass}, \binits{M.}},
\oauthor{\bsnm{Pong}, \binits{V.H.}},
\oauthor{\bsnm{Powell}, \binits{T.}},
\oauthor{\bsnm{Power}, \binits{A.}},
\oauthor{\bsnm{Power}, \binits{B.}},
\oauthor{\bsnm{Proehl}, \binits{E.}},
\oauthor{\bsnm{Puri}, \binits{R.}},
\oauthor{\bsnm{Radford}, \binits{A.}},
\oauthor{\bsnm{Rae}, \binits{J.}},
\oauthor{\bsnm{Ramesh}, \binits{A.}},
\oauthor{\bsnm{Raymond}, \binits{C.}},
\oauthor{\bsnm{Real}, \binits{F.}},
\oauthor{\bsnm{Rimbach}, \binits{K.}},
\oauthor{\bsnm{Ross}, \binits{C.}},
\oauthor{\bsnm{Rotsted}, \binits{B.}},
\oauthor{\bsnm{Roussez}, \binits{H.}},
\oauthor{\bsnm{Ryder}, \binits{N.}},
\oauthor{\bsnm{Saltarelli}, \binits{M.}},
\oauthor{\bsnm{Sanders}, \binits{T.}},
\oauthor{\bsnm{Santurkar}, \binits{S.}},
\oauthor{\bsnm{Sastry}, \binits{G.}},
\oauthor{\bsnm{Schmidt}, \binits{H.}},
\oauthor{\bsnm{Schnurr}, \binits{D.}},
\oauthor{\bsnm{Schulman}, \binits{J.}},
\oauthor{\bsnm{Selsam}, \binits{D.}},
\oauthor{\bsnm{Sheppard}, \binits{K.}},
\oauthor{\bsnm{Sherbakov}, \binits{T.}},
\oauthor{\bsnm{Shieh}, \binits{J.}},
\oauthor{\bsnm{Shoker}, \binits{S.}},
\oauthor{\bsnm{Shyam}, \binits{P.}},
\oauthor{\bsnm{Sidor}, \binits{S.}},
\oauthor{\bsnm{Sigler}, \binits{E.}},
\oauthor{\bsnm{Simens}, \binits{M.}},
\oauthor{\bsnm{Sitkin}, \binits{J.}},
\oauthor{\bsnm{Slama}, \binits{K.}},
\oauthor{\bsnm{Sohl}, \binits{I.}},
\oauthor{\bsnm{Sokolowsky}, \binits{B.}},
\oauthor{\bsnm{Song}, \binits{Y.}},
\oauthor{\bsnm{Staudacher}, \binits{N.}},
\oauthor{\bsnm{Such}, \binits{F.P.}},
\oauthor{\bsnm{Summers}, \binits{N.}},
\oauthor{\bsnm{Sutskever}, \binits{I.}},
\oauthor{\bsnm{Tang}, \binits{J.}},
\oauthor{\bsnm{Tezak}, \binits{N.}},
\oauthor{\bsnm{Thompson}, \binits{M.B.}},
\oauthor{\bsnm{Tillet}, \binits{P.}},
\oauthor{\bsnm{Tootoonchian}, \binits{A.}},
\oauthor{\bsnm{Tseng}, \binits{E.}},
\oauthor{\bsnm{Tuggle}, \binits{P.}},
\oauthor{\bsnm{Turley}, \binits{N.}},
\oauthor{\bsnm{Tworek}, \binits{J.}},
\oauthor{\bsnm{Uribe}, \binits{J.F.C.}},
\oauthor{\bsnm{Vallone}, \binits{A.}},
\oauthor{\bsnm{Vijayvergiya}, \binits{A.}},
\oauthor{\bsnm{Voss}, \binits{C.}},
\oauthor{\bsnm{Wainwright}, \binits{C.}},
\oauthor{\bsnm{Wang}, \binits{J.J.}},
\oauthor{\bsnm{Wang}, \binits{A.}},
\oauthor{\bsnm{Wang}, \binits{B.}},
\oauthor{\bsnm{Ward}, \binits{J.}},
\oauthor{\bsnm{Wei}, \binits{J.}},
\oauthor{\bsnm{Weinmann}, \binits{C.}},
\oauthor{\bsnm{Welihinda}, \binits{A.}},
\oauthor{\bsnm{Welinder}, \binits{P.}},
\oauthor{\bsnm{Weng}, \binits{J.}},
\oauthor{\bsnm{Weng}, \binits{L.}},
\oauthor{\bsnm{Wiethoff}, \binits{M.}},
\oauthor{\bsnm{Willner}, \binits{D.}},
\oauthor{\bsnm{Winter}, \binits{C.}},
\oauthor{\bsnm{Wolrich}, \binits{S.}},
\oauthor{\bsnm{Wong}, \binits{H.}},
\oauthor{\bsnm{Workman}, \binits{L.}},
\oauthor{\bsnm{Wu}, \binits{S.}},
\oauthor{\bsnm{Wu}, \binits{J.}},
\oauthor{\bsnm{Wu}, \binits{M.}},
\oauthor{\bsnm{Xiao}, \binits{K.}},
\oauthor{\bsnm{Xu}, \binits{T.}},
\oauthor{\bsnm{Yoo}, \binits{S.}},
\oauthor{\bsnm{Yu}, \binits{K.}},
\oauthor{\bsnm{Yuan}, \binits{Q.}},
\oauthor{\bsnm{Zaremba}, \binits{W.}},
\oauthor{\bsnm{Zellers}, \binits{R.}},
\oauthor{\bsnm{Zhang}, \binits{C.}},
\oauthor{\bsnm{Zhang}, \binits{M.}},
\oauthor{\bsnm{Zhao}, \binits{S.}},
\oauthor{\bsnm{Zheng}, \binits{T.}},
\oauthor{\bsnm{Zhuang}, \binits{J.}},
\oauthor{\bsnm{Zhuk}, \binits{W.}},
\oauthor{\bsnm{Zoph}, \binits{B.}}:
GPT-4 Technical Report
(2024).
\doiurl{10.48550/arXiv.2303.08774} .
\url{https://arxiv.org/abs/2303.08774}
\end{botherref}
\endbibitem

\bibitem[\protect\citeauthoryear{Ofori-Boateng et~al.}{2024}]{Ofori}
\begin{botherref}
\oauthor{\bsnm{Ofori-Boateng}, \binits{R.}},
\oauthor{\bsnm{Aceves-Martins}, \binits{M.}},
\oauthor{\bsnm{Wiratunga}, \binits{N.}},
\oauthor{\bsnm{Moreno-García}, \binits{C.}}:
Towards the automation of systematic reviews using natural language processing, machine learning, and deep learning: a comprehensive review.
Artificial Intelligence Review
\textbf{57}
(2024)
\doiurl{10.1007/s10462-024-10844-w}
\end{botherref}
\endbibitem

\bibitem[\protect\citeauthoryear{Probst et~al.}{2019}]{RN205}
\begin{barticle}
\bauthor{\bsnm{Probst}, \binits{P.}},
\bauthor{\bsnm{Boulesteix}, \binits{A.-L.}},
\bauthor{\bsnm{Bischl}, \binits{B.}}:
\batitle{Tunability: importance of hyperparameters of machine learning algorithms}.
\bjtitle{Journal of Machine Learning Research}
\bvolume{20}(\bissue{1}),
\bfpage{1934}--\blpage{1965}
(\byear{2019})
\doiurl{10.48550/arXiv.1802.09596}
\end{barticle}
\endbibitem

\bibitem[\protect\citeauthoryear{Pintas et~al.}{2021}]{RN190}
\begin{barticle}
\bauthor{\bsnm{Pintas}, \binits{J.T.}},
\bauthor{\bsnm{Fernandes}, \binits{L.A.F.}},
\bauthor{\bsnm{Garcia}, \binits{A.C.B.}}:
\batitle{Feature selection methods for text classification: a systematic literature review}.
\bjtitle{Artificial Intelligence Review}
\bvolume{54}(\bissue{8}),
\bfpage{6149}--\blpage{6200}
(\byear{2021})
\doiurl{10.1007/s10462-021-09970-6}
\end{barticle}
\endbibitem

\bibitem[\protect\citeauthoryear{Suzuki et~al.}{2023}]{SUZUKI2023103194}
\begin{barticle}
\bauthor{\bsnm{Suzuki}, \binits{M.}},
\bauthor{\bsnm{Sakaji}, \binits{H.}},
\bauthor{\bsnm{Hirano}, \binits{M.}},
\bauthor{\bsnm{Izumi}, \binits{K.}}:
\batitle{Constructing and analyzing domain-specific language model for financial text mining}.
\bjtitle{Information Processing \& Management}
\bvolume{60}(\bissue{2}),
\bfpage{103194}
(\byear{2023})
\doiurl{10.1016/j.ipm.2022.103194}
\end{barticle}
\endbibitem

\bibitem[\protect\citeauthoryear{Tu et~al.}{2024}]{tu2024chatlog}
\begin{barticle}
\bauthor{\bsnm{Tu}, \binits{S.}},
\bauthor{\bsnm{Li}, \binits{C.}},
\bauthor{\bsnm{Yu}, \binits{J.}},
\bauthor{\bsnm{Wang}, \binits{X.}},
\bauthor{\bsnm{Hou}, \binits{L.}},
\bauthor{\bsnm{Li}, \binits{J.}}:
\batitle{{ChatLog: Carefully Evaluating the Evolution of ChatGPT Across Time}}.
\bjtitle{arXiv pre-print server}
(\byear{2024})
\doiurl{10.48550/arXiv.2002.10957}
\end{barticle}
\endbibitem

\bibitem[\protect\citeauthoryear{Wang et~al.}{2020}]{RN197}
\begin{barticle}
\bauthor{\bsnm{Wang}, \binits{W.}},
\bauthor{\bsnm{Wei}, \binits{F.}},
\bauthor{\bsnm{Dong}, \binits{L.}},
\bauthor{\bsnm{Bao}, \binits{H.}},
\bauthor{\bsnm{Yang}, \binits{N.}},
\bauthor{\bsnm{Zhou}, \binits{M.}}:
\batitle{{MiniLM: Deep Self-Attention Distillation for Task-Agnostic Compression of Pre-Trained Transformers}}.
\bjtitle{Advances in Neural Information Processing Systems}
\bvolume{33},
\bfpage{5776}--\blpage{5788}
(\byear{2020})
\doiurl{10.48550/arXiv.2002.10957}
\end{barticle}
\endbibitem

\bibitem[\protect\citeauthoryear{Yu et~al.}{2022}]{RN202}
\begin{barticle}
\bauthor{\bsnm{Yu}, \binits{Y.-X.}},
\bauthor{\bsnm{Gong}, \binits{H.-P.}},
\bauthor{\bsnm{Liu}, \binits{H.-C.}},
\bauthor{\bsnm{Mou}, \binits{X.}}:
\batitle{Knowledge representation and reasoning using fuzzy {P}etri nets: a literature review and bibliometric analysis}.
\bjtitle{Artificial Intelligence Review}
(\byear{2022})
\doiurl{10.1007/s10462-022-10312-3}
\end{barticle}
\endbibitem

\end{thebibliography}

\end{document}